\documentclass[reprint,twocolumn,aps,prl,amsmath,amssymb,floatfix,superscriptaddress,longbibliography]{revtex4-2}
\usepackage{float}
\usepackage[T1]{fontenc}
\usepackage{pifont} 
\usepackage{dcolumn}
\usepackage{enumitem}
\usepackage{makecell, adjustbox}
\usepackage{graphicx}
\graphicspath{ {images/}, {figs/} }

\usepackage{bm,dsfont}
\usepackage{physics,comment}
\usepackage[colorlinks=true,linktoc=all,linkcolor=blue, citecolor=blue, filecolor=blue, urlcolor=blue]{hyperref}
\usepackage[capitalise]{cleveref} 
\usepackage{xcolor,color}
\usepackage[caption=false]{subfig}
\usepackage{orcidlink}
\makeatletter
\newcommand{\colorcaption}[2][]{%
  \begingroup%
  \renewcommand{\@caption@fignum@sep}{ (color online). }%
  \caption[#1]{#2}%
  \endgroup%
} \makeatother
\newcommand{\tb}[1]{\textbf{#1}}

\def\red{\textcolor{red}}

\def\be{\begin{equation}} \def\ee{\end{equation}}
\def\bea{\begin{eqnarray}} \def\eea{\end{eqnarray}}
\def\nn{\nonumber}

\definecolor{profreplycolor}{RGB}{255,0,255}

\begin{document}

\title{Hilbert space connectivity in non-Hermitian many-body systems: emergent scale-dependent amplification and constraint-induced skin localization}

\author{Zichang Hao\,\orcidlink{0009-0009-3397-3443}}
\affiliation{Department of Physics, National University of Singapore, Singapore 117542}
\author{Wen-Tan Xue\,\orcidlink{0000-0001-7823-9888}}
\affiliation{Department of Physics, National University of Singapore, Singapore 117542}
\author{Ching Hua Lee\,\orcidlink{0000-0003-0690-3238}}
\email{{phylch@nus.edu.sg}}
\affiliation{Department of Physics, National University of Singapore, Singapore 117542}

\date{\today}
\begin{abstract}
Various exotic many-body phenomena such as quantum scars and fractons have been linked to Hilbert space fragmentation. In this work, we find that in non-Hermitian settings, Hilbert space connectivity has an even more universal and fundamental influence, tightly controlling the nature of spectral amplification and state localization. Far more complicated than real-space lattices,  non-Hermitian many-body Hilbert space graphs not only possess intricate competing amplification channels, but also global feedback loops connecting remote Fock states related by particle symmetry. These features lead to amplification behavior with unconventional scaling and localization properties. Particle occupation constraints can furthermore remove selected Hilbert space pathways, leading to robust unipolar and asymmetric bipolar skin localization in otherwise reciprocal processes.  
These results extend beyond simple interacting bosonic models and establish Hilbert space connectivity as a versatile control knob for many-body non-Hermitian critical transitions.

\end{abstract}
\maketitle

\noindent\red{\emph{Introduction.}}---
In non-Hermitian single-particle systems, it is already well established that non-reciprocal hopping can drive extensive state accumulation at the boundaries, in a phenomenon known as the non-Hermitian skin effect (NHSE)~\cite{HatanoNelson1997Vortex,lee2016anomalous,yao2018non,lee2019anatomy,song2019non,Lee2019Hybrid,OkumaSato2023Review,
ashida2020non,bergholtz2021exceptional,gong2018topological,kawabata2019symmetry,okuma2020topological,yang2022designing,sanahal2025gauge,hu2025many,wang2025nonlinear,zhao2025two,zhang2026harmonic,Yang2026Reversing,Okuma2026steady,Longhi2026Mpemba,qin2024kinked,Kokkinakis2026Defect,Zhu2021Delocalization,shen2025non,lin2023topological}. 
When multiple skin channels compete, the accumulation can become system size (scale)-dependent, leading to the critical non-Hermitian skin effect (cNHSE)~\cite{li2020critical,yokomizo2021scaling,qin2023universal,Qin2025Many,Liu2022Helical,rafi2022unconventional,guo2021exact,cheng2025stochasticity,zhang2024observation}. A basic open problem is how this picture changes in interacting many-body systems, where momentum-space band descriptions such as the generalized Brillouin zone GBZ~\cite{yao2018edge,yokomizo2019non,YokomizoMurakami2020PTEP,Yang2020AuxGBZ,li2025phase,meng2025generalized,yuan2026non} no longer apply.

In many-body settings, a more natural viewpoint is the connectivity of Hilbert space, which is determined by interactions and constraints~\cite{shen2024enhanced,liu2025quantized,moudgalya2022hilbert,moudgalya2022quantum}. In Hermitian systems, such Hilbert space structure can qualitatively shape state dynamics, for instance through fragmentation and constrained thermalization~\cite{hudomal2020quantum,moudgalya2022hilbert,moudgalya2022quantum,yang2020hilbert,sala2020ergodicity,mukherjee2021minimal,hahn2021information,Moudgalya2021}.
Meanwhile, non-Hermitian many-body systems have been shown to exhibit many-body localization~\cite{hamazaki2019non,suthar2022non,li2023non,zhai2020many,Wang2026longrange,Mak2024Statics,Liu2023ergodicity,hao2026interacting}, various many-body skin effects~\cite{shen2022non,alsallom2022fate,lee2021many,yoshida2024non,shimomura2024general,gliozzi2024many,qin2024occupation,Hamanaka2024,koh2025interacting,Qin2025Many,Li2023dynamicgauge,hu2025many,yang2025non,kim2024collective,qin2025dynamical,shen2025observation}, and unconventional many-body correlations and entanglement~\cite{mu2020emergent,nakagawa2018non,yu2024non,Turkeshi2023,He2026graph,kawabata2023entanglement,lee2022exceptional,chang2020entanglement,Garcia2022,Hyart2022,Han2023Nonclassicality,liu2025non,xue2026topologically}, but how Hilbert space connectivity interplays with the NHSE in many-body systems remains largely unexplored.

A useful precedent for this viewpoint arises from fractional quantum Hall (FQH) physics, where local clustering rules and generalized Pauli principles organize the allowed many-body occupation configurations ~\cite{haldane1991fractional,neupert2011fractional,yang2012model,lee2014lattice,zhu2013minimal,yang2017generalized,xie2021fractional}. 
In Moore--Read and Read--Rezayi states, for instance, these rules have profound links to conformal-field-theory fusion rules and non-Abelian quasiparticle statistics \cite{moore1991nonabelions,read1999beyond,nayak2008non}. Jack-polynomial descriptions make this Hilbert space organization explicit: a root occupation pattern, together with squeezing and clustering rules, defines the accessible Hilbert space of model FQH states \cite{bernevig2008model,bernevig2008generalized,hansson2017quantum}. 
Indeed, clustering constraints not only determine which configurations are allowed, but also how the Hilbert space is connected~\cite{trugman1985exact,lee2004mott,seidel2005incompressible,lee2015geometric,xu2026fractional}. 
Here we ask: What new roles can analogous clustering rules play when the allowed Hilbert space pathways are non-Hermitian and directional?

In this work, we show that in non-Hermitian settings, interaction-engineered Hilbert space connectivity naturally reshapes the competition between multiple coupled non-reciprocal amplification channels. Due to this amplification, the resultant dynamics turn out completely different from the clustering behavior of occupation-constrained Hermitian systems~\cite{TaoThouless,RezayiHaldane1994,lee2004mott,seidel2005incompressible,BergholtzKarlhede,xu2026fractional}. This will be demonstrated in detail below via an interacting boson model with quenched single-particle kinetics, even though most results generalize to more sophisticated models with further hoppings and higher-order interactions [see also the End Matter].

\begin{figure*}[htbp]
    \centering
    \includegraphics[width=0.95\linewidth]{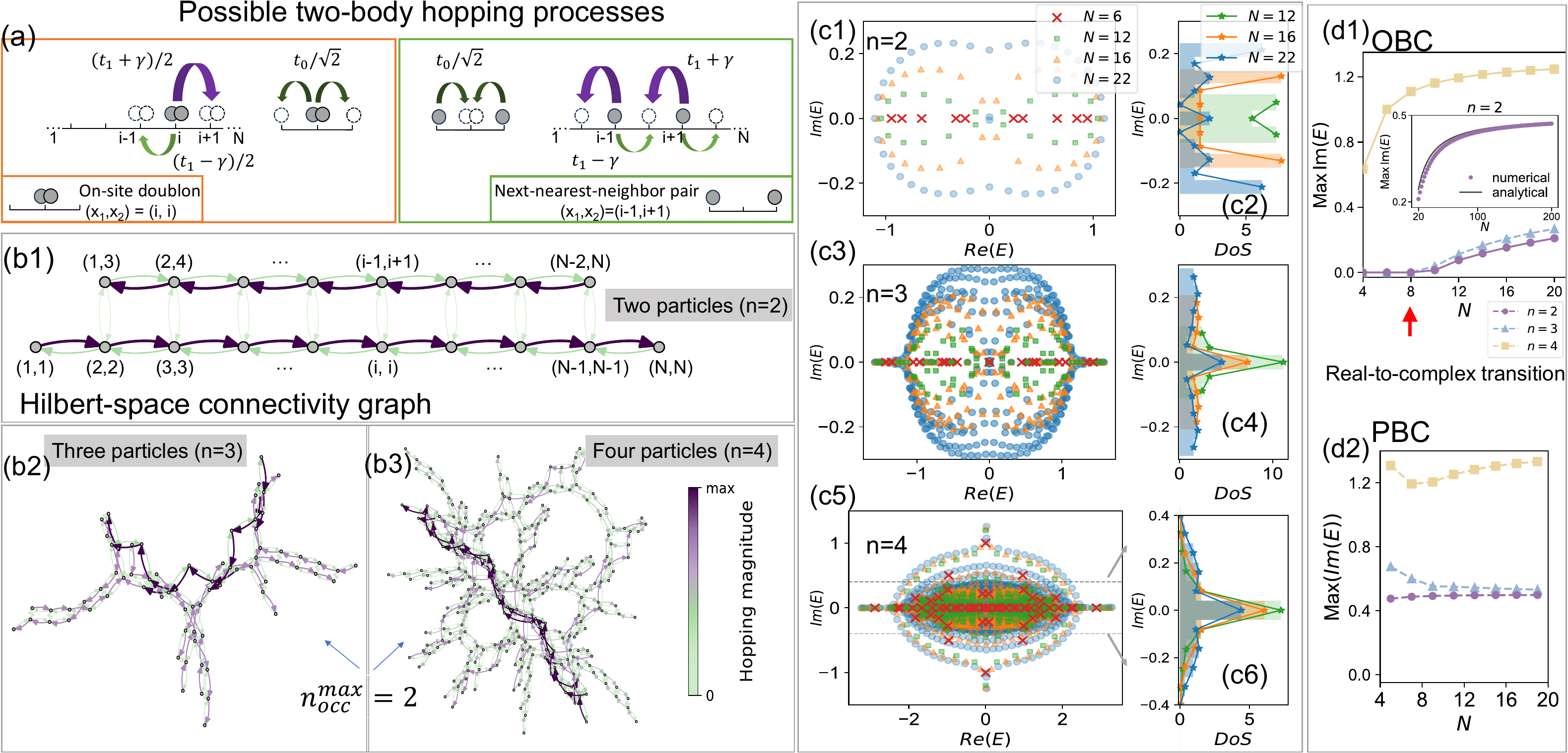}
    \caption{\textbf{Hilbert space connectivity structure and scaling-induced spectral transitions for the interacting boson model of \cref{eq:model}.} 
    \textbf{(a)} Illustration of two representative occupation configurations and their allowed hoppings: (Left) two particles occupying the same site $(i,i)$ i.e., a doublon, and (Right) two particles separated by one intervening site $(i-1,i+1)$. Asymmetric hoppings (purple/light green) keep the particles within the same type of configuration, while symmetric hoppings $t_0/\sqrt{2}$ (dark-green arrows) convert between these configurations. Non-Hermiticity arises from $\gamma \neq 0$. 
    \textbf{(b)} Hilbert space connectivity graphs of \cref{eq:model} for two-, three-, and four-particle systems at system size $N=10$ under OBCs~\cite{evenGexp}, highlighting the antagonistic-coupled chains as the basic structural motif. 
    \textbf{(c)} Eigenenergy spectra $E$ and the density of states (DOS) resolved by  ${\operatorname{Im}}(E)$ for two-( c1-c2), three- (c3-c4), and four-particle (c5-c6) systems, with salient dependence on the system size $N=6, 12, 16, 22$. 
    \textbf{(d)} Scaling behavior of $\max \operatorname{Im}(E)$ versus $N$ for OBCs~[(d1)] and PBCs~[(d2)]. OBCs are characterized by pronounced size-driven real-to-complex spectral transitions, whereas no comparable transition is observed under PBCs. 
    Inset of (d1): comparison between the two-particle ($n=2$) numerical results and the large-$N$ GBZ prediction~[\cref{eq:two-boson-asymp}].
    Parameter settings: $t_1 = 0.58$, $\gamma = 0.25$, $t_0 = 0.01$. All results obey a maximum on-site occupancy $n_{\mathrm{occ}}^{\max}=2$.
        }
    \label{fig:Fig1}
\end{figure*}

\noindent\red{\emph{Scaling-induced real-to-complex transition and Hilbert space reorganization based on particle number}.}---To illustrate emergent interaction-induced non-Hermitian physics, we first consider a minimal 1D bosonic model with three qualitatively distinct types of correlated two-body hoppings [\cref{fig:Fig1}(a)]. 
The Hamiltonian is 
\begin{equation}
\begin{aligned}
    H &=\sum_{i=1}^{N} \frac{t_{1} + \gamma}{2}c_{i+1}^{\dagger}c_{i}c_{i+1}^{\dagger}c_{i} + \frac{t_{1} - \gamma}{2}c_{i}^{\dagger}c_{i+1}c_{i}^{\dagger}c_{i+1} \\
    &+(t_{1} - \gamma)c_{i+2}^{\dagger}c_{i+1}c_{i}^{\dagger}c_{i-1} + (t_{1} + \gamma)c_{i-1}^{\dagger}c_{i}c_{i+1}^{\dagger}c_{i+2}\\
    &+  \frac{t_0}{\sqrt{2}} c_{i+1}^{\dagger}c_{i}c_{i-1}^{\dagger}c_{i} + \frac{t_{0}}{\sqrt{2}}c_{i}^{\dagger}c_{i-1}c_i^{\dagger}c_{i+1},
\end{aligned}
\label{eq:model}
\end{equation}
where $c_i (c_i^{\dagger})$ is the bosonic annihilation (creation) operator on site $i$ and $N$ is the system size. Despite its simplicity, $H$ already shares much of the behavior of more general models involving further and more complicated interactions, as elaborated in Supp. Sec.~SIV~\cite{mysupp}  and the End Matter.

The first line in Eq.~\eqref{eq:model} describes on-site pairs (doublons) jointly hopping to a neighboring site with unbalanced amplitudes $(t_1\pm \gamma)/2$ (purple, light green). The second line also describes unbalanced correlated hoppings $t_1\mp \gamma$ (light green, purple), but for next-nearest-neighbors separated by one intervening site. The third line couples these two configurations, i.e., on-site pairs and next-nearest-neighbor pairs through oppositely-directed hoppings $t_0/\sqrt{2}$ (dark green) in or out of a doublon. 
Here the prefactors $1/2$ and $1/\sqrt{2}$ are introduced to offset multiple counting from particle statistics, such as to obtain the cleanest effective model possible. 
Notably, this Hamiltonian does not act on single-particle states, guaranteeing that any observed phenomenon arises exclusively from particle interactions rather than single-body non-Hermitian skin physics~\cite{hatano1996localization}. 

As a warm-up, we begin with the two-boson case under open boundary conditions (OBCs), and show that the interactions are already sufficient for inducing a real-to-complex spectral transition i.e., emergent amplification controlled by system size.   
This is most easily understood by examining the accessible Hilbert space, which takes the form of two coupled nonreciprocal chains [\cref{fig:Fig1}(b1)] indexed by $(i,i)$ and $(i-1,i+1)$, since the hoppings only connect on-site doublon or next-nearest-neighbor pair configurations [\cref{fig:Fig1}(a)].  
In this picture, the dynamics on the Hilbert space graph mimics that on an effective single-particle hopping model, with nodes representing many-body Fock state configurations and links encoding the allowed correlated hoppings [see Supp. Sec.~SI~\cite{mysupp}]. 
For this simplest 2-boson example, the effective model is none other than that of the cNHSE ladder~\cite{li2020critical,yokomizo2021scaling,qin2023universal}: for sufficiently large $N$ or strong interchain couplings $t_0$, the  nonreciprocal dynamics form feedback loops circulating around the ladder, leading to a complex spectrum; this picture is moot under PBCs, as detailed in  Supp. Sec.~SVI~\cite{mysupp}. 
As shown in \cref{fig:Fig1}(c1), a real-to-complex transition indeed occurs as $N$ is increased -- rigorously, it approaches the asymptotic amplification rate $2\gamma$ as $1/N$:
\begin{equation}
\max \operatorname{Im}(E)\simeq 2\gamma-(2t_1/N)\ln((t_1+\gamma)/t_0),
\label{eq:two-boson-asymp}
\end{equation}
as derived in Supp. Sec.~SII~\cite{mysupp} and numerically verified in the inset of \cref{fig:Fig1}(d1).

We now move on to focus on $n=3$ or more particles, where the simple correspondence with cNHSE breaks down, and genuinely interacting phenomena emerge. First consider a maximal on-site occupancy of $n_{\mathrm{occ}}^{\max}=2$. 
Although certain parts of the Hilbert space graph still inherit competing Hatano--Nelson ladder structures from $n=2$ subspaces, they serve only as local building blocks picked out by interaction constraints [see \cref{fig:Fig1}(b2–b3), with links colored by hopping strength]. Links between these ladders correspond to special accumulation spots described later, and bosonic permutation symmetry ``glue'' distant configurations to create additional loops.

With more particles, scaling-induced real-to-complex transitions generally persist, but with caveats. For $n=3$, the transition is still distinct, as shown in \cref{fig:Fig1}(c3) and \cref{fig:Fig1}(d1). For $n=4$ [\cref{fig:Fig1}(c5)], however, a cleanly real spectrum only exists at very small $N$, though the overall complexification trend remains evident.  
This universal tendency towards larger Im$(E)$ can be seen from the broadening of the density of states (DOS) along ${\rm Im}(E)$ in \cref{fig:Fig1}(c6), as well as from the growth of ${\rm max}[{\rm Im}(E)]$ in \cref{fig:Fig1}(d1). See Supp.~Sec.~SV~\cite{mysupp} for more detailed IPR and entanglement entropy characterizations.

The qualitative difference between having $n=3$ and $n=4$ bosons can be understood from the connectivity of the Hilbert space graph. For $n=3$, the graph retains a predominantly branch-like structure [\cref{fig:Fig1}(b2)], where each branch forms a cNHSE-like feedback loop  
[\cref{fig:Fig1}(b1)]. 
Hence the effect of having $n=3$ particles is to couple multiple $(n=2)$-like building blocks, inheriting their real-to-complex transition [\cref{fig:Fig1}(c3-c4)].

With $n=4$ bosons, the feedback loops are no longer confined within individual branches: Instead, due to greater Fock space redundancy from bosonic symmetry, some branches connect to form large, system-spanning loops in the Hilbert space graph [\cref{fig:Fig1}(b3)]. An arbitrary state can therefore circulate across multiple branches and undergo repeated nonreciprocal amplification along these new global feedback pathways, generating Im$(E)>0$ eigenenergies even at small system sizes [\cref{fig:Fig1}(c5-c6)]. Despite being physically under OBCs, such global paths are reminiscent of periodic boundary condition (PBC) loops -- with similar amplification rates Im$(E)$ for both PBCs and OBCs [yellow in \cref{fig:Fig1}(c,d)]. 
Since the circumferences of these permutation symmetry-induced loops scale with $N$, a stochastic estimate [Supp. Sec.~SIII~\cite{mysupp}] yields, for 4 or more bosons,
\begin{equation}
\max\operatorname{Im}E\simeq M_\infty-A/N+O(N^{-2})
\label{maxImE4}
\end{equation}
where $M_\infty,A$ are system-specific constants, generalizing \cref{eq:two-boson-asymp} with excellent numerical agreement [Fig.~S4].

\begin{figure}[htbp]
    \centering
    \includegraphics[width=\linewidth]{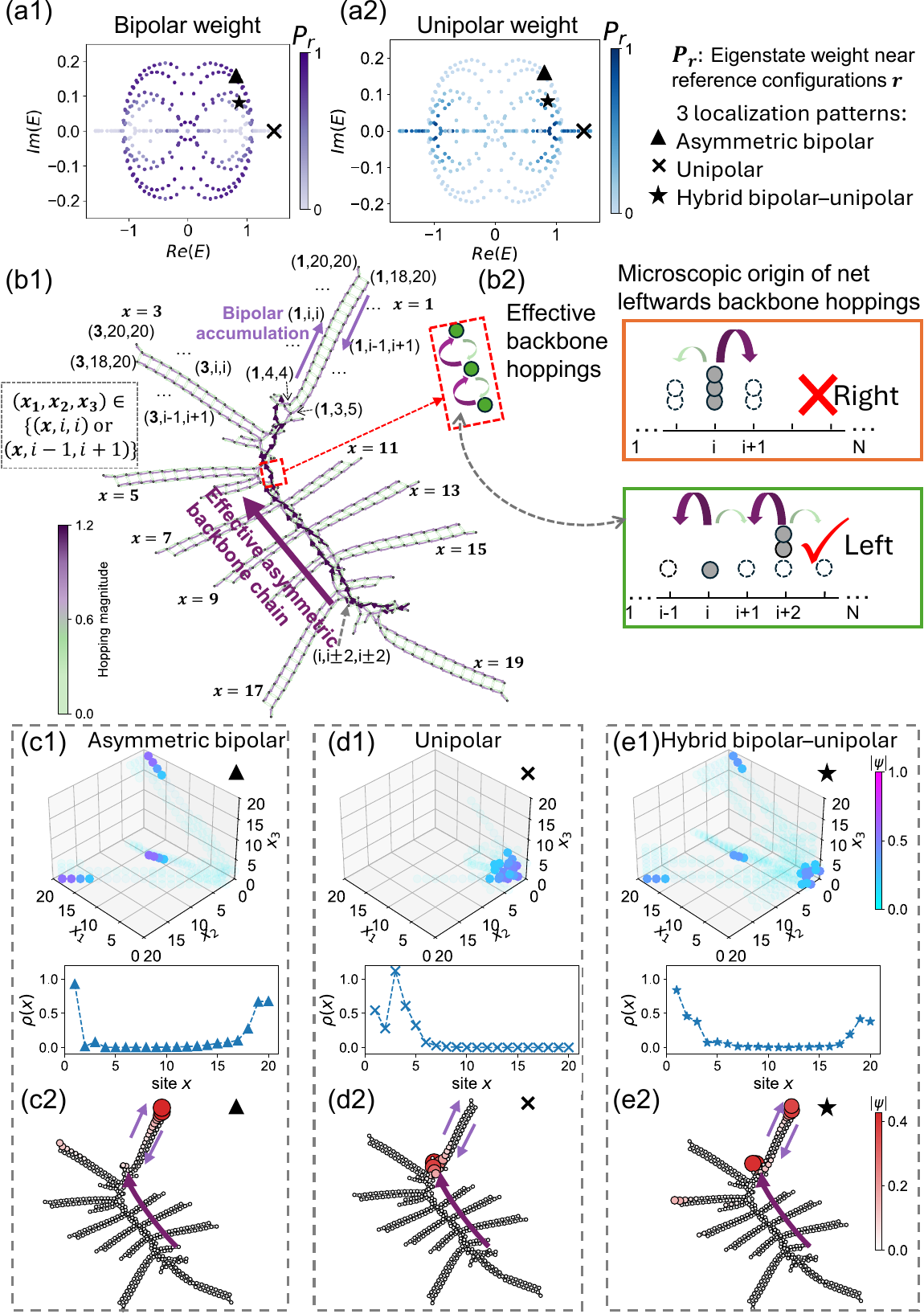}
    \caption{\textbf{Unipolar and asymmetric bipolar localization from effective asymmetric backbone skin pumping induced by the occupation constraint $n_{\mathrm{occ}}^{\max}=2$.} 
    \textbf{(a1-a2)} Energy spectrum of \cref{eq:model} colored by (a1) asymmetric bipolar and (a2) unipolar weight $P_{\bm r}$~[\cref{eq:local-prob}]. 
    \textbf{(b1)} Hilbert space connectivity graph of \cref{eq:model}, with a central $(i,i\pm 2,i\pm 2)$ backbone connected to multiple branches. Each branch corresponds to parallel chains of configurations $(x,i,i)$ and $(x,i-1,i+1)$, which contains a boson at $x$ and two other bosons centered at $i$. The purple arrow denotes the constraint-induced directional hopping bias along the backbone. 
    \tb{(b2)} The rightward pumping process (Top) is inactive due to forbidden triple occupancy, leaving a net leftward-favoring pumping process (Bottom) along the backbone.
    \textbf{(c1,d1,e1)} (Top) Amplitude distributions $|\psi|$ of representative localized eigenstates marked in (a1–a2) in $(x_1,x_2,x_3)$ configuration space;
    (Bottom) their averaged particle density distribution $\rho(x)=\langle \psi|c_x^{\dagger}c_x|\psi\rangle$ in the Hilbert space. \textbf{(c2,d2,e2)} The same eigenstates as in (c1–e1), shown on the Hilbert space connectivity graph. 
    The parameters are set to $t_0=10^{-3}$, $t_1=0.58$, system size $N=20$, and $\gamma=0.25$.
    }
    \label{fig:Fig2}
\end{figure}

\noindent\red{\emph{Effective asymmetric pumping and edge localization from occupation constraints.}}---
Here, we further discuss how the on-site occupancy constraint $n_{\mathrm{occ}}^{\max}<n$ can shape the Hilbert space and induce asymmetric skin pumping, 
sticking to our minimal example with three bosons ($n=3$) and $n_{\mathrm{occ}}^{\max}=2$. 
The central finding is that removing Fock states with more than $n_{\mathrm{occ}}^{\max}$ particles gives rise to directed amplification ``bottlenecks'' in the Hilbert space, leading to strong edge localization beyond simple NHSE: \textbf{unipolar}, \textbf{asymmetric bipolar}, or the coexistence of both (\textbf{hybrid}). 
To characterize the localization pattern for an eigenstate $\psi$, we define $P_{\bm r}(\psi)$, the weight of $\psi$ within a cutoff distance $\delta_{\bm r}$ from a given configuration $\bm r$:
\begin{equation}
    P_{\bm r}(\psi) = \sum_{\|(x_1,x_2,x_3)-\bm r\|_2\le \delta_{\bm r}} |\psi(x_1, x_2, x_3)|^2.
    \label{eq:local-prob}
\end{equation}
To quantify the extent of unipolar and asymmetric bipolar localization, we consider $\bm r = (1,1,1)$ and $\bm r=(1,N,N)$ respectively, both with $\delta_{\bm r}=5$~\footnote{The conclusions are generally insensitive to the cutoff distance, as shown in Supp. Sec.~SIV~\cite{mysupp}.} 
From \cref{fig:Fig2}(a1) and (a2), eigenstates with larger $|\text{Im}(E)|$ tend to be more bipolar-localized (dark purple), while those with larger $|\text{Re}(E)|$ are likely more unipolar (dark blue); states which exhibit both localizations are termed hybrid bipolar-unipolar states. Representative states, marked by $\blacktriangle$, \scalebox{1.2}{$\bm{\times}$}, and $\bigstar$ in \cref{fig:Fig2}(a1,a2), are plotted in the three-boson configuration space $(x_1,x_2,x_3)$ in \cref{fig:Fig2}(c1--e1). Unipolar states have all 3 particles localized at the $x=1$ end, while asymmetric bipolar states have 2 particles at one end and 1 particle at the other.

To explain the emergence of these state configurations, we examine the Hilbert space connectivity graph [\cref{fig:Fig2}(b1)] in detail. 
Overall, it consists of a central chain-like backbone with multiple outward branches attached. 
The backbone corresponds to configurations where all three particles are close to each other  [\cref{fig:Fig2}(b2), bottom], i.e., $(i, i\pm 2, i\pm 2)$, since triple occupancy is excluded by $n_{\mathrm{occ}}^{\max}=2$. 
The branches are indexed by $x$, each containing two parallel chains with configurations $ (x, i, i)$ and $(x, i-1, i+1)$ i.e., where two particles form a bound pair (doublon or next-nearest neighbor), while the other one is freely located at $x$.

The central backbone exhibits a strongly leftwards net directionality [purple arrow in \cref{fig:Fig2}(b1)] even though the first two lines of Eq.~\eqref{eq:model} suggest equal and opposite hopping asymmetry ratios $\frac{t_1\pm \gamma}{t_1\mp \gamma}$.
The reason is elucidated in \cref{fig:Fig2}(b2): of these two antagonistic microscopic processes, only the net leftwards process in the green box can occur, since it acts on valid backbone configurations $(i, i\pm 2, i\pm 2)$. There, left hoppings (purple arrow) $c_{i-1}^\dagger c_i c_{i+1}^\dagger c_{i+2}$ occur with stronger amplitudes $t_1+\gamma$ than for the reverse process (green arrow), whose amplitude is $t_1-\gamma$. 
By contrast, the net rightwards hopping mode [orange box of \cref{fig:Fig2}(b2)] is inactive because of forbidden triply occupied states $(i,i,i)$. In effect, the occupancy constraint $n_{\mathrm{occ}}^{\max}=2$ has thus given rise to an effective net leftward pumping along the backbone. 

This occupancy constraint-induced Hilbert space pumping is the key towards understanding the emergence of unipolar and asymmetric bipolar localization routes.  
As illustrated in \cref{fig:Fig2}(b1), it first universally drives states leftwards towards the head of the backbone indicated by the deep-purple arrow, where the $x=1$ branch joins. 
States that continue to be funneled into the top branch 
are bipolar-localized around $(1,N,N)$ [\cref{fig:Fig2}(c2)].    
But since the allowed transitions among $(1,i,i)$ and $(1,i-1,i+1)$ configurations in the branch form two coupled subchains with opposite hopping biases, analogous to \cref{fig:Fig1}(b1), some states also remain unipolar-localized near the branch junction $(1,1,1)$ [\cref{fig:Fig2}(d2)].   
Unipolar localization occurs in the blue spectral region of \cref{fig:Fig2}(a2) and is consistent with a point-gap-induced NHSE~\cite{okuma2020topological}, as the spectrum contracts toward the real axis upon switching from PBCs to OBCs [see Supp. Sec.~SVI~\cite{mysupp}]. Hybrid states with substantial weight in both localization patterns give the mixed profiles shown in \cref{fig:Fig2}(e1,e2).

\begin{figure}[t]
    \centering
    \includegraphics[width=\linewidth]{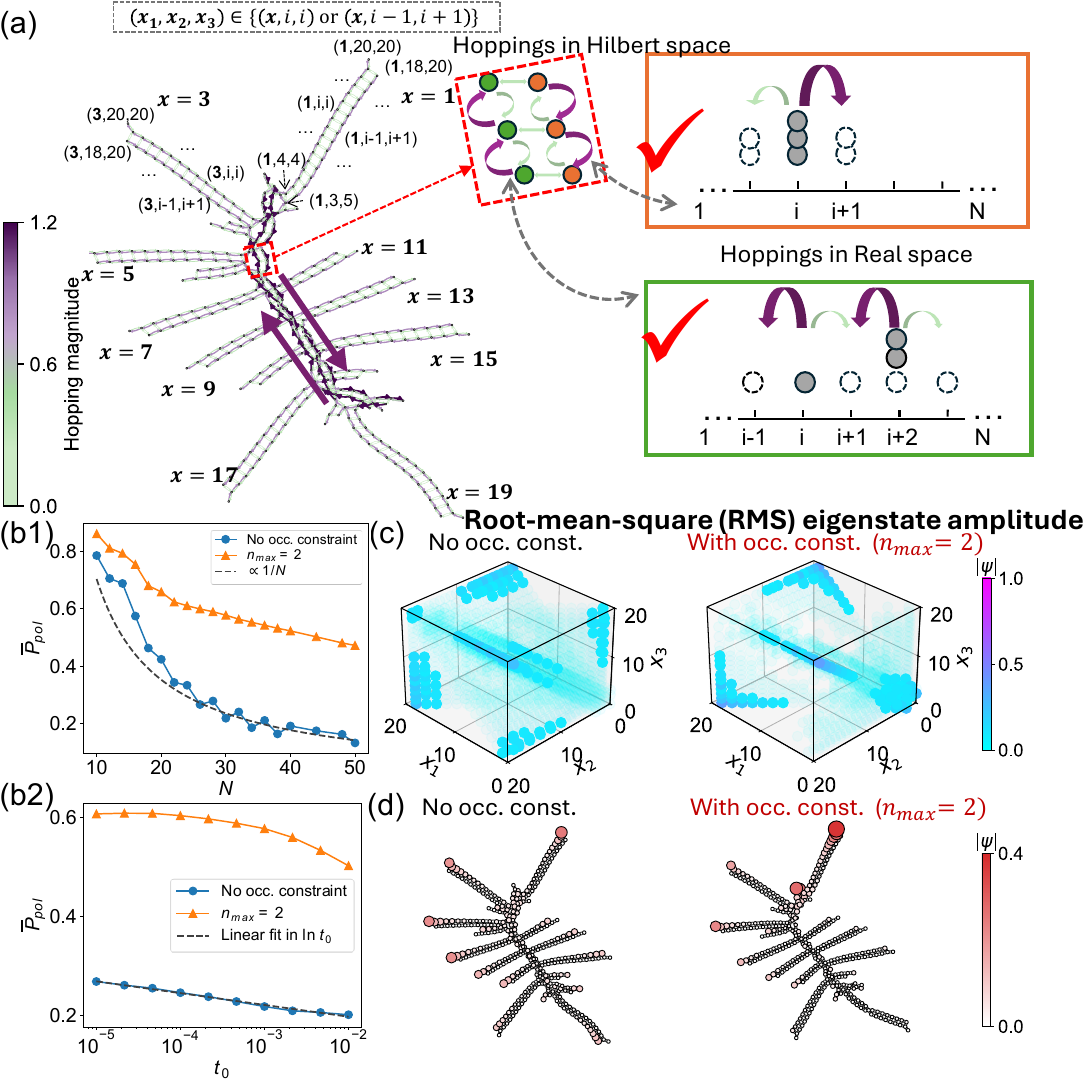}
    \caption{
    Removing the occupancy constraint ($n_{\mathrm{occ}}^{\max} = 2$) restores the reverse boson-enhanced backbone process and qualitatively changes the scaling of corner polarization.
    Results for a three-particle system without an occupancy constraint under OBCs.
    \textbf{(a)} (Left) Hilbert space connectivity graph.
    (Right) the correlated-hopping process forbidden in \cref{fig:Fig2} becomes allowed once the occupancy constraint is removed.
    \textbf{(b)} Average polarization weight $\bar{P}_{\text{pol}}$ [\cref{eq:P-ave-corner}], shown with and without the occupancy constraint, under varying system sizes $N$ and couplings $t_0$. 
    \textbf{(b1)} $\bar{P}_{\text{pol}}$ versus system size $N$ at fixed $t_0=10^{-3}$, with an approximately $1/N$ decay in the unconstrained system~(blue), compared to much slower decay with  $n_{\mathrm{occ}}^{\max} = 2$ (orange). 
    \textbf{(b2)} $\bar{P}_{\text{pol}}$ versus coupling $t_0$ at fixed $N=20$.
    \textbf{(c-d)} Root-mean-square (RMS) eigenstate amplitude with and without the occupancy constraint, shown in configuration space [(c)] and on the Hilbert space connectivity graph [(d)].
    Unless otherwise noted, parameters are $t_0=10^{-3}$, $t_1=0.58$, $\gamma=0.25$, and $N=20$.
      }
    \label{fig:Fig3}
\end{figure}

\noindent\red{\emph{Relaxation of the occupancy constraint.}}--- To isolate the effects of the occupancy constraint on the unipolar/bipolar localization of the eigenstates, we study the above system without enforcing this constraint.
This leaves the microscopic hoppings in \cref{eq:model} unchanged, but enlarges the accessible Hilbert space by restoring configurations of the form $(i,i,i)$ [\cref{fig:Fig3}(a), right]. These restored states reactivate the reverse boson-enhanced backbone process, thereby weakening the net pumping responsible for both unipolar and asymmetric bipolar localization [\cref{fig:Fig3}(a), left].

We quantify the system-wide strength of unipolar/bipolar localization by averaging the eigenstate weight  $P_{\bm r}$ [\cref{eq:local-prob}] over all eigenstates $\psi_n$ and all corners, defined as the average polarization weight,
\begin{equation}
\bar{P}_{\text{pol}}=\frac{1}{\mathcal D}\sum_{n=1}^{\mathcal D}\sum_{\bm r} P_{\bm r}(\psi_n),
\label{eq:P-ave-corner}
\end{equation}
where $\mathcal D$ is the total number of eigenstates, and $\bm{r} \in \{(1,1,1),(1,1,N),(1,N,N),(N,N,N)\}$.

The quantity $\bar{P}_{\mathrm{pol}}$ measures, on average, how strongly eigenstates display unipolar/bipolar localization.
\cref{fig:Fig3}(b) shows that removing the occupancy constraint significantly suppresses $\bar P_{\text{pol}}$ across system sizes $N$ and couplings $t_0$ (orange $\to$ blue), demonstrating that the constraint is \emph{necessary} for robust unipolar/bipolar localization. It is further supported by \cref{fig:Fig3}(c,d), where we compare root-mean-square (RMS) eigenstate amplitude $\sqrt{\langle j|\rho|j\rangle}$ \cite{RMSexp} across configurations $j$ on the microcanonical density matrix $\rho$, as well as center-of-mass measures in Supp. Sec.~SVII~\cite{mysupp}. 
For the unconstrained case, the observed
$\bar P_{\mathrm{pol}}\propto 1/N$ scaling follows naturally from the scale-free nature of critical skin coupling: along a branch of length $\ell$, the
wave function decays exponentially, $\psi(r)\sim e^{-r/\xi}$, with a
scale-free localization length $\xi\propto\ell$~\cite{li2020critical,qin2023universal}. Hence, a fixed corner cutoff
$\delta_r=5$ captures only a fraction
$\sim\delta_r/\xi\propto 1/\ell\propto 1/N$ of their weight, where system size $N\approx\ell$ .
The logarithmic $t_0$ dependence also follows from the exponential decay of the
skin modes.
Imposing the occupancy constraint enhances both unipolar and bipolar localization [\cref{fig:Fig3}(c), left to right], with the probability weight concentrating more strongly near the upper branch [\cref{fig:Fig3}(d), left to right].  

Taken together, robust unipolar and asymmetric bipolar localization requires (i) asymmetric physical hoppings, i.e, $\gamma\neq 0$, (ii) occupation-dependent bosonic factors, originating from $c_i\ket{n_i}=\sqrt{n_i}\ket{n_i-1}$, that enhance correlated hoppings on clustered configurations, and (iii) a constraint that inactivates the reverse enhanced process. All these key amplification and localization mechanisms from particle statistics and constraints persist in more general interacting non-Hermitian hopping models, as detailed in the End Matter.

\noindent\red{\emph{Discussion.}}---
We have demonstrated that in interacting non-Hermitian systems, many-body NHSE behavior is chiefly controlled by Hilbert space connectivity rather than real-space hopping topology. 
While the dynamics behind a pair of interacting particles simply reduces to scale-dependent cNHSE, 
having more particles exposes the physics to 
non-trivial effects from on-site occupancy constraints and particle statistics. 
Occupancy constraints remove configurations that would otherwise host competing channels, leading to emergent selective asymmetric pumping and robust unipolar and asymmetric bipolar localization. Particle statistics furthermore alter the global structure of the Hilbert space profoundly, giving rise to feedback loops with no spatial analog.

\begin{acknowledgments}
\red{\emph{Acknowledgments.}}---Some of the exact diagonalization simulations have been performed using QuSpin~\cite{phillip2017quspin,phillip2019quspin}. ZH would like to thank Ruizhe Shen and Yi Qin for their helpful discussions. We acknowledge support from the Ministry of Education, Singapore (Award No.~MOE-T2EP50224-0007).
\end{acknowledgments}

\bibliography{references}
\clearpage
\section*{End Matter}

\twocolumngrid

\subsection*{Appendix A: Generality beyond the minimal interacting model}

The phenomena presented in the main text -- real-to-complex transitions and constraint-induced localization -- extend far beyond our bosonic interaction model, as elaborated in Supp. Sec.~SIV~\cite{mysupp}, where more general parameter sets, occupation constraints, hopping ranges and even interaction mechanisms are examined. 
In particular, similar scaling behavior persists when the correlated hopping range is increased from the present case of $d=2$ to $d=3$ and $d=4$, and when the interaction is replaced by a three-body term. 
These examples indicate that the essential ingredient is the coexistence of distinct asymmetric hopping channels coupled through the many-body Hilbert space, rather than a special fine-tuned geometry.

As such, even when the Hilbert space no longer possesses an easily decipherable structure [see \cref{fig:end-generality} for visualizations], the physics remains qualitatively similar:
for $n>2$-body interactions, such feedback loops generally occur in irregular connectivity graphs that do not possess any likeness to regular non-interacting lattices. 
We further examine whether the scaling-induced real-to-complex spectral behavior found in the main text persists in more general interacting non-Hermitian models. Specifically, we consider two broader classes of interacting models: (i) two-body correlated hopping with different hopping distances $d$, and (ii) Hamiltonians involving three-body correlated hopping processes. Although Hilbert-space connectivity graphs for these models may be more intricate, their spectra display the same qualitative behavior: as the system size increases, an initially nearly real spectrum develops pronounced broadening along $\operatorname{Im}(E)$. 

\subsection{Longer-range two-body correlated hoppings}
\begin{figure}[htbp]
    \centering
    \includegraphics[width=\linewidth]{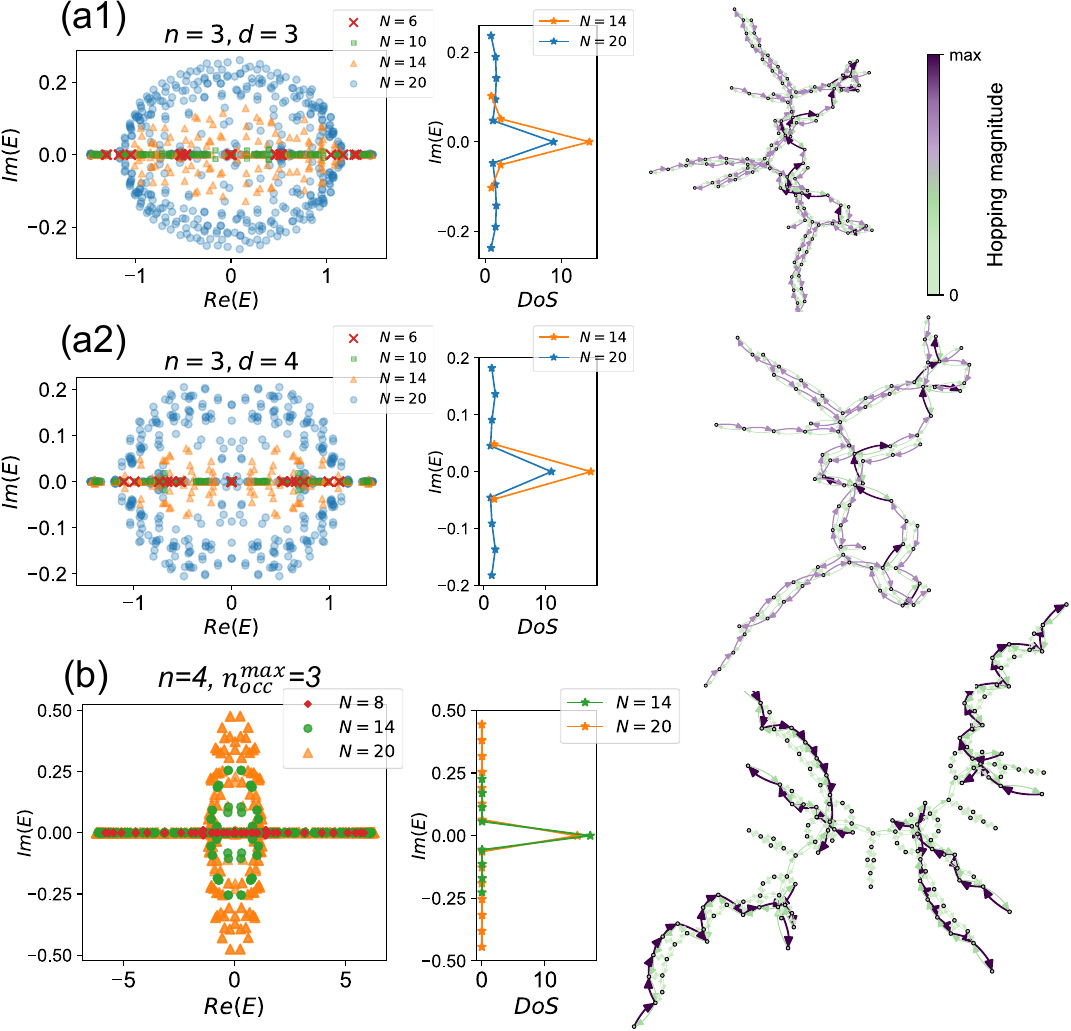}
    \caption{\textbf{Scaling-induced real-to-complex spectral broadening and Hilbert-space connectivity in generalized interacting models.}
    \textbf{(a1,a2)} Longer-range two-body correlated-hopping model [\cref{eq:end-long-range}] for three particles ($n=3$) with hopping distances $d=3$ and $d=4$, respectively, under the occupancy constraint $n_{\mathrm{occ}}^{\max}=2$.
    \textbf{(b)} Three-body correlated-hopping model [\cref{eq:end-three-body}] for four particles ($n=4$) with $n_{\mathrm{occ}}^{\max}=3$.
    For each case, the eigenvalue spectra are shown for several system sizes $N$, together with the corresponding $\operatorname{Im}(E)$-resolved density of states and a representative Hilbert-space connectivity graph.
    Throughout, $t_1=t_2=0.58$, $t_0=10^{-2}$, and $\gamma=0.25$, with OBCs imposed.
    }
    \label{fig:end-generality}
\end{figure}

We generalize the correlated-hopping distance from $d=2$ in the main-text model to an arbitrary value $d$. The generalized Hamiltonian with correlated hopping distance $d$ is given by
\begin{widetext}
\begin{equation}
\begin{aligned}
H &= \sum_{i=1}^{N} \frac{t_0}{\sqrt{2}} c_{i+1}^{\dagger}c_{i}c_{i+1-d}^{\dagger}c_{i} + \frac{t_{0}}{\sqrt{2}}c_{i}^{\dagger}c_{i+1-d}c_i^{\dagger}c_{i+1} +\frac{t_{1} + \gamma}{2}c_{i+1}^{\dagger}c_{i}c_{i+1}^{\dagger}c_{i} + \frac{t_{1} - \gamma}{2}c_{i}^{\dagger}c_{i+1}c_{i}^{\dagger}c_{i+1} \\
&+(t_{2} - \gamma)c_{i+2}^{\dagger}c_{i+1}c_{i+2-d}^{\dagger}c_{i+1-d} + (t_{2} + \gamma)c_{i+1-d}^{\dagger}c_{i+2-d}c_{i+1}^{\dagger}c_{i+2}.
\end{aligned}
\label{eq:end-long-range}
\end{equation}
\end{widetext}
As shown in \cref{fig:end-generality}(a1-a2), a real-to-complex transition persists across $d=3,4$. In particular, the eigenvalue distribution exhibits significant broadening along the imaginary axis. This is quantitatively confirmed by the density of states (DOS) resolved by the imaginary part of the eigenvalues, which shows a clear spread along the $\operatorname{Im}(E)$ axis—indicating the robustness of the real-to-complex transition. 

The persistence of this behavior can be understood from the Hilbert-space connectivity [\cref{fig:end-generality}(a1-a2), right]: changing $d$ only changes the real-space diameters of the clusters, while preserving locally embedded, oppositely biased Hatano--Nelson-like channels inherited from the $n=2$ subspaces. Their coupling generates the same nonreciprocal feedback mechanism responsible for the size-dependent spectral complexification.

\subsection{Three-body correlated hoppings}
We next consider the three-body correlated-hopping Hamiltonian, for which a generalized Hamiltonian is given by 
\begin{widetext}
\begin{equation}
\begin{aligned}
    H_{3-body} &= \sum_{i=1}^{N} (t_1+\gamma) c_{i+1}^{\dagger}c_i c_{i+1}^{\dagger}c_i c_{i+1}^{\dagger}c_i + (t_1-\gamma)c_{i}^{\dagger}c_{i+1} c_{i}^{\dagger}c_{i+1} c_{i}^{\dagger}c_{i+1} \\
    & + (t_2-\gamma) c_{i}^{\dagger}c_{i-1} c_{i+1}^{\dagger}c_i c_{i+2}^{\dagger}c_{i+1} + (t_2+\gamma)c_{i+1}^{\dagger}c_{i+2} c_{i}^{\dagger}c_{i+1} c_{i-1}^{\dagger}c_{i} \\
    &+ t_0 c_{i-1}^{\dagger}c_{i} c_{i}^{\dagger}c_i c_{i+1}^{\dagger}c_{i} + t_0c_{i}^{\dagger}c_{i+1} c_{i}^{\dagger}c_{i} c_{i}^{\dagger}c_{i-1}.
    \label{eq:end-three-body}
\end{aligned}
\end{equation}
\end{widetext}

For $n=3$, the three-body processes generate the oppositely biased cluster-translation channels $(i,i,i)\rightleftharpoons(i+1,i+1,i+1)$ and $(i-1,i,i+1)\rightleftharpoons(i,i+1,i+2)$, which are coupled through the $t_0$ processes. Together, they form a cNHSE-like coupled-chain motif at the level of three-particle clusters. For $n=4$, the additional particle interconnects multiple such local motifs, so that a simple global ladder mapping no longer applies. Nevertheless, these oppositely biased channels remain the local building blocks of the Hilbert-space connectivity~[see \cref{fig:end-generality}(b), right].
Their $t_0$-mediated coupling creates nonreciprocal feedback pathways within the larger Hilbert-space connectivity graph, thereby retaining the same size-dependent spectral broadening shown in \cref{fig:end-generality}(b).

\section{Physical implementation prospects}

A direct physical consequence in a system with non-Hermitian many-body correlated hoppings/interactions is that the system size controls both the spectral amplification/decay rate, measured by $\max \operatorname{Im}(E)$, and the accompanying real-space accumulation pattern. Because both signatures are accessible in few-body sectors, they are natural targets in the digital quantum simulation of non-Hermitian dynamics~\cite{shen2025observation,zhang2025observation,koh2025interacting,shen2026simulating,shen2026observation,shen2025robust}. Such dynamics can be implemented on rapidly advancing quantum hardware~\cite{ladd2010quantum,preskill2018quantum,alexeev2021quantum}, leveraging variational circuit optimization~\cite{mcardle2019variational,chen2023high,koh2022stabilizing,shen2025observation,koh2024realization}, non-unitary postselection~\cite{chen2023high,shen2025observation,lin2021real}, and state-of-the-art error-mitigation techniques~\cite{temme2017error,endo2018practical,shen2025circuit}. The amplification rate $\max \operatorname{Im}(E)$ can be measured following the procedure outlined in Sec.~SVIII of~\cite{mysupp}, which directly probes complex spectra. More broadly, the results highlight Hilbert space structure as a practical design principle for non-Hermitian many-body criticality, complementary to conventional real-space engineering.

\clearpage
\newpage

\setcounter{equation}{0}
\setcounter{figure}{0}
\setcounter{table}{0}
\setcounter{section}{0}
\setcounter{subsection}{0}

\renewcommand{\theequation}{S\arabic{equation}}
\renewcommand{\thefigure}{S\arabic{figure}}
\renewcommand{\thetable}{S\arabic{table}}
\renewcommand{\thesection}{S\arabic{section}}

\renewcommand{\theHequation}{supp.equation.\arabic{equation}}
\renewcommand{\theHfigure}{supp.figure.\arabic{figure}}
\renewcommand{\theHtable}{supp.table.\arabic{table}}
\renewcommand{\theHsection}{supp.section.\arabic{section}}
\renewcommand{\theHsubsection}{supp.subsection.\arabic{subsection}}

\onecolumngrid
\setcounter{page}{1} 
\renewcommand{\thepage}{S\arabic{page}}
\flushbottom

\appendix
\begin{center}
\textbf{\large Supplemental Material for “Hilbert space connectivity in non-Hermitian many-body systems: emergent scale-dependent amplification and constraint-induced skin localization” }
\end{center}

\begin{center}
 {\small Zichang Hao$^{1}$, Wen-Tan Xue$^{1}$, and Ching Hua Lee$^{1}$ }  
\end{center}
\begin{center}
{\sl \footnotesize

$^{1}$Department of Physics, National University of Singapore, Singapore 117542
}
\end{center}
\begin{quote}
	{\small
    This Supplemental Material is organized as follows: \\ 
    (SI) Model and mapping of the interacting two-particle model onto an effective lattice (single-body model).\\
    (SII) Derivation of the size-dependent two-particle spectrum. \\
    (SIII) Estimation of scaling behaviour of four-particle system.\\
    (SIV) Robustness and generality of scaling-induced real-to-complex transition.\\  
    (SV) Eigenstate signatures of real-to-complex transition from entanglement entropy and inverse participation
    ratio.\\
    (SVI) Boundary-condition dependence of real-to-complex transition and Hilbert space connectivity. \\
    (SVII) Constraint-induced Hilbert space pumping and unipolar/asymmetric bipolar localization. \\
    (SVIII) Possible proposal to measure the scaling-induced maximum imaginary eigenenergy ($\text{maxIm}(E)$) on a quantum processor.
	}   
\end{quote}

\section{SI. Model and mapping of the interacting two-particle model onto an effective lattice}

We begin by identifying the minimal Hilbert space structure underlying scaling-induced real-to-complex transition:
in the two-particle sector, the interacting model maps onto two coupled nonreciprocal chains with opposite hopping
biases.

\begin{figure}[htbp]
    \centering
    \includegraphics[width=0.6\linewidth]{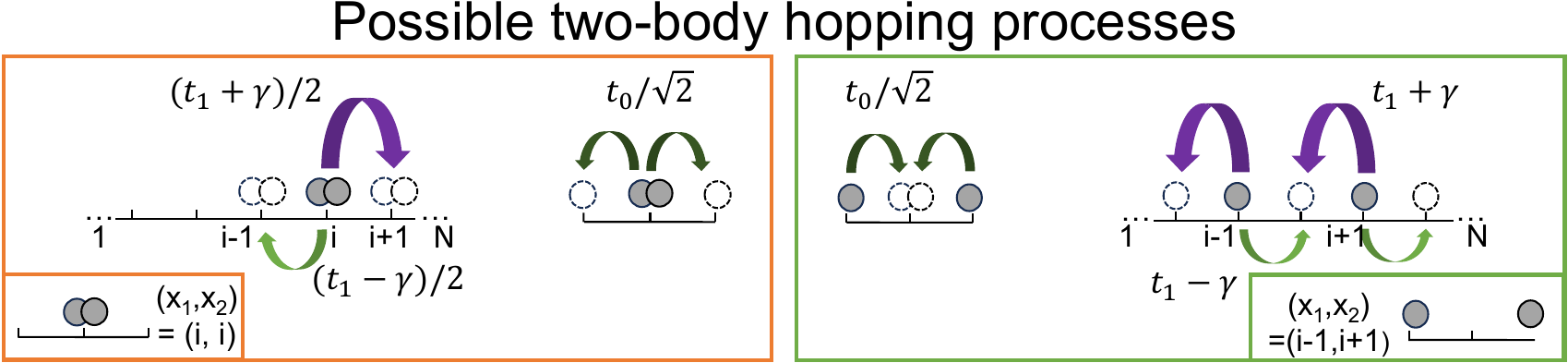}
    \caption{Illustration of two representative occupation configurations in \cref{eq:supp-model}: (Left) two particles occupying the same site $(i,i)$, and (Right) two particles separated by one intervening site $(i-1,i+1)$. Within each configuration, asymmetric intra-configuration hoppings (orange and green arrows) occur, while the two configurations are connected by a reciprocal inter-configuration hopping (dark-green arrows).}
    \label{fig:supp-hopping-skech}
\end{figure}

\paragraph{Model}
As introduced in the main text, we consider a one-dimensional bosonic model
\begin{equation}
\begin{aligned}
    H &=\sum_{i=1}^{N} \frac{t_{1} + \gamma}{2}c_{i+1}^{\dagger}c_{i}c_{i+1}^{\dagger}c_{i} + \frac{t_{1} - \gamma}{2}c_{i}^{\dagger}c_{i+1}c_{i}^{\dagger}c_{i+1} \\
    &+(t_{1} - \gamma)c_{i+2}^{\dagger}c_{i+1}c_{i}^{\dagger}c_{i-1} + (t_{1} + \gamma)c_{i-1}^{\dagger}c_{i}c_{i+1}^{\dagger}c_{i+2}\\
    &+  \frac{t_0}{\sqrt{2}} c_{i+1}^{\dagger}c_{i}c_{i-1}^{\dagger}c_{i} + \frac{t_{0}}{\sqrt{2}}c_{i}^{\dagger}c_{i-1}c_i^{\dagger}c_{i+1},
\end{aligned}
\label{eq:supp-model}
\end{equation} 
where $c_i (c_i^{\dagger})$ is the bosonic annihilation (creation) operator at the $i$-th site and $N$ is the system size. The Hamiltonian consists of three types of correlated two-body hoppings
(illustrated in \cref{fig:supp-hopping-skech}): (i) hopping of an on-site pair, (ii) hopping of a next-nearest-neighbor
pair (separated by one site), and (iii) interconversion between these two pair configurations controlled by $t_0$.
 
\paragraph{Two-particle mapping to the effective lattice}
Here, we show how our interacting model can be mapped onto a single-body model. If we consider a two-particle system $n=2$, we can plot the hoppings in real space $(x_1,x_2)$, where $x_1, x_2$ denote the locations of two particles separately. Then we find that it effectively forms two coupled chains with opposite hopping biases~\cite{li2020critical, yokomizo2021scaling}.
\begin{figure}[H]
    \centering
    \includegraphics[width=0.8\linewidth]{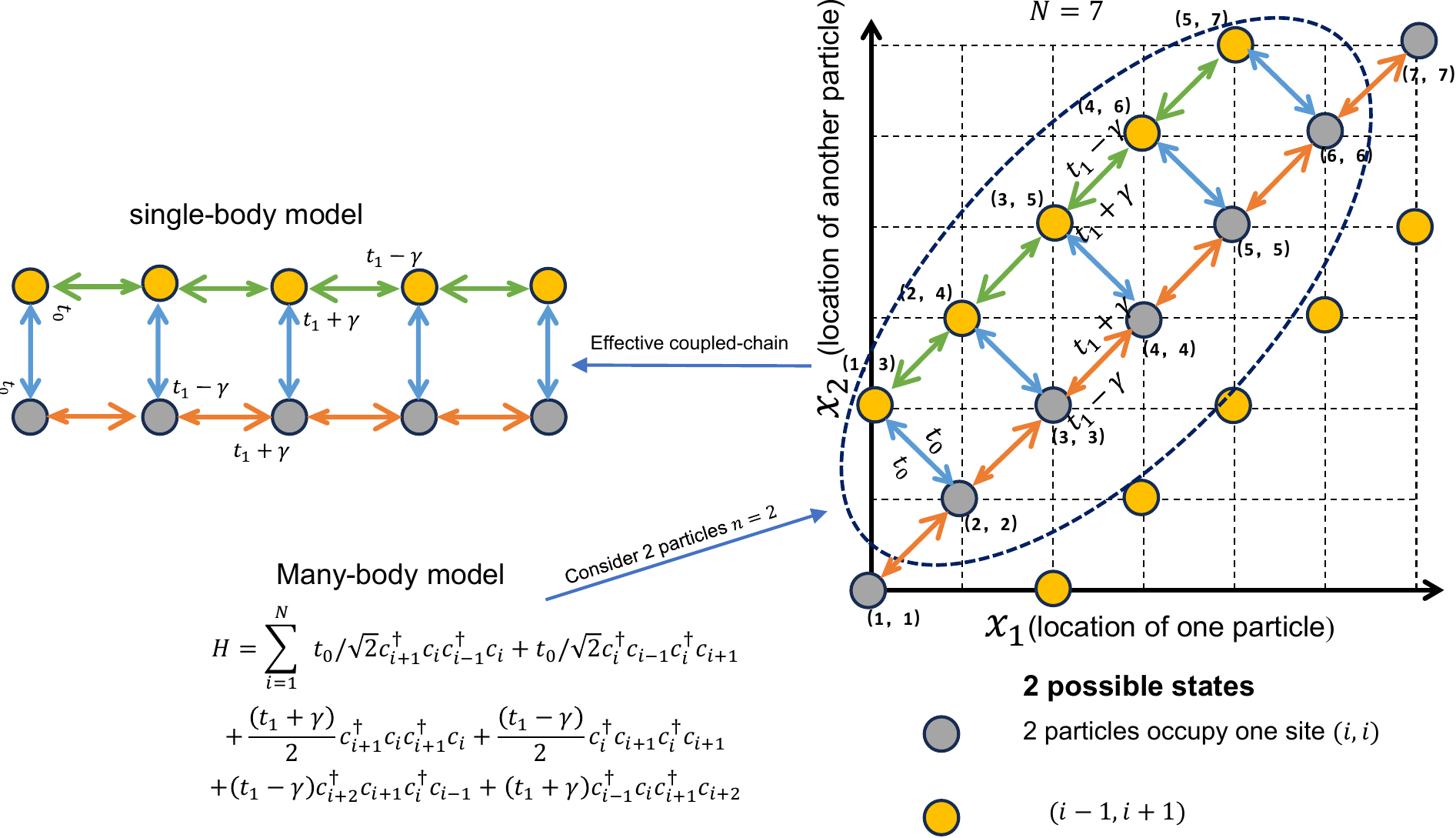}
    \caption{
    Mapping of the interacting many-body model [\cref{eq:supp-model}, with total particle number $n=2$] onto an effective single-particle two-chain model in Hilbert space.
    }
    \label{fig:supp-2body-sector-hop}
\end{figure}
This mapping gives the 2-particle sector an effective coupled-chain description. Based on it, we show in Sec.~SII how to analytically derive its size-dependent spectrum.

\section{SII. Derivation of scaling-dependent analytic spectra for two-particle systems}

\begin{figure}[H]
    \centering
    \includegraphics[width=0.8\linewidth]{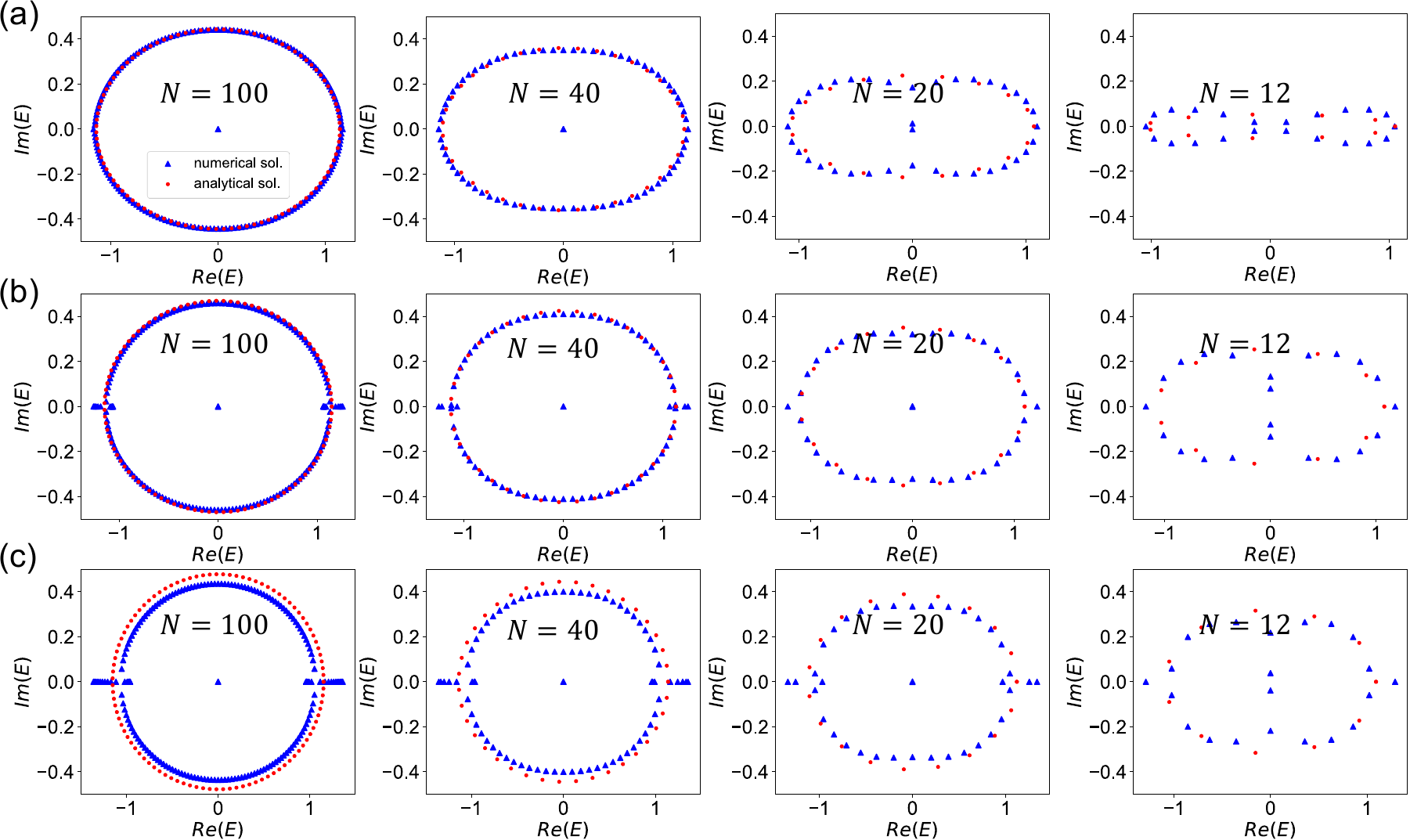}
    \caption{
    Numerical verification ('numerical sol.') of the eigenvalue spectra of
    the active sector of the two-particle ($n=2$) model
    in \cref{eq:supp-model-gbz},
    compared with the approximate analytic prediction
    ('analytical sol.') in \cref{eq:E_ana}
    for different system sizes $N$ and values of $t_0$.
    Parameters: $t_1=t_2=0.58, \gamma=0.25.$
    \textbf{(a)} $t_0=0.01$
    \textbf{(b)} $t_0=0.1$
    \textbf{(c)} $t_0=0.2$.
    The analytical spectra agree well with the numerical results,
    particularly for small $t_0$ and large $N$.
    }
    \label{fig:supp-verification-GBZ}
\end{figure}

In this section, we derive the scaling-dependent eigenvalue spectrum for the generalized two-particle system governed by the Hamiltonian:
\bea
H&=&\sum_{i=1}^{N} \Bigl[\frac{t_0}{\sqrt{2}} c_{i+1}^{\dagger}c_{i}c_{i-1}^{\dagger}c_{i} + \frac{t_{0}}{\sqrt{2}}c_{i}^{\dagger}c_{i-1}c_i^{\dagger}c_{i+1} \nn\\
&+&\frac{t_{1} + \gamma}{2}c_{i+1}^{\dagger}c_{i}c_{i+1}^{\dagger}c_{i} + \frac{t_{1} - \gamma}{2}c_{i}^{\dagger}c_{i+1}c_{i}^{\dagger}c_{i+1}\nn \\
&+&(t_{2} - \gamma)c_{i+2}^{\dagger}c_{i+1}c_{i}^{\dagger}c_{i-1} + (t_{2} + \gamma)c_{i-1}^{\dagger}c_{i}c_{i+1}^{\dagger}c_{i+2}\Bigr],
\label{eq:supp-model-gbz}
\eea
which corresponds to the hopping distance-$(d=2)$ case of the general model introduced in \cref{eq:supp-gen-model-1}.
Under OBC, hopping terms involving sites outside $1,\ldots,N$ are omitted.
As illustrated in \cref{fig:supp-2body-sector-hop},
the dynamically connected sector spanned by $(i,i)$ (chain I) and $(i-1,i+1)$ (chain II)
can be mapped onto two coupled chains in Hilbert space.
For the asymptotic bulk analysis below, we omit the two terminal configurations $(x_1,x_2)=(1,1)$ and $(N,N)$; the resulting approximation is tested against the exact finite-chain spectrum in \cref{fig:supp-verification-GBZ}.
Each reduced chain has $N-2$ sites; in the leading large-$N$ approximation below, we replace this length by $N$.
The remaining OBC system has a translationally invariant bulk described by the Bloch Hamiltonian:
\be
H(k)=\begin{pmatrix}
(t_1-\gamma)e^{ik}+(t_1+\gamma)e^{-ik} & t_0\\
t_0 & (t_2+\gamma)e^{ik}+(t_2-\gamma)e^{-ik}
\end{pmatrix}.
\ee
In the following, by deriving the effective scaling-dependent generalized Brillouin zone (GBZ) and the corresponding approximate OBC spectral envelope associated with $H(k)$, we can effectively predict the nontrivial two-particle spectrum of the interacting system in Eq.~\eqref{eq:supp-model-gbz}.

Considering real parameters $t_1,t_2>\gamma>0$ and $t_0>0$, the uncoupled two chains exhibit non-Hermitian skin effect in opposite directions.
For a weak-coupling approximation, we retain only the inter-chain couplings at the two edges and neglect those in the bulk.
Under this simplification, we assume the following ansatz for the wavefunctions:
\be
\psi^\mathrm{I}(x)\sim
\alpha^x \text{\quad for chain I,\quad\quad and }
\psi^{\mathrm{II}}(x)\sim \beta^x
\text{\quad for chain II}.
\label{eq:GBZbulkEq}
\ee
The corresponding bulk equations for the two chains are given by:
$\left\{
\begin{aligned}
   (t_1-\gamma)\alpha+(t_1+\gamma)/\alpha &= E \\
   (t_2+\gamma)\beta+(t_2-\gamma)/\beta &= E,
\end{aligned}
\right. $
which yield four roots $\alpha_{1,2}$ and $\beta_{1,2}$ satisfying:
\be
\left\{
\begin{aligned}
   (t_1-\gamma)\alpha_{1,2}+(t_1+\gamma)/\alpha_{1,2}
   &= E,\quad
   \text{with }\alpha_1\alpha_2=\frac{t_1+\gamma}{t_1-\gamma}\\
   (t_2+\gamma)\beta_{1,2}+(t_2-\gamma)/\beta_{1,2}
   &= E,\quad
   \text{with }\beta_1\beta_2=\frac{t_2-\gamma}{t_2+\gamma}.
\end{aligned}
\right.
\label{eq:abroots}
\ee
By ordering the roots as $|\alpha_1|\leq|\alpha_2|$ and $|\beta_1|\leq|\beta_2|$, we obtain $|\alpha_2|>1$ and $|\beta_1|<1$, which will be used in subsequent derivations.
For distinct roots, the wavefunctions at the two chains, $\psi^\mathrm{I}(x)$ and $\psi^\mathrm{II}(x)$, can then be written as the superpositions:
\be
\left\{
\begin{aligned}
   \psi^\mathrm{I}(x)=c_1\alpha_1^x+c_2\alpha_2^x,
   & \quad\text{for chain I}\\
   \psi^\mathrm{II}(x)=d_1\beta_1^x+d_2\beta_2^x,
   & \quad\text{for chain II},
\end{aligned}
\label{eq:GBZsuper}
\right.
\ee
where $c_{1,2}$ and $d_{1,2}$ are the corresponding superposition coefficients.

To determine the values of $\alpha_{1,2}$ and $\beta_{1,2}$, we substitute $\psi^\mathrm{I}(x)$ and $\psi^\mathrm{II}(x)$ from \cref{eq:GBZsuper} into the following four boundary conditions of the edge-coupled approximation at the two edges:
\be
\left\{
\begin{aligned}
   t_0\psi^\mathrm{II}(1)+(t_1-\gamma)\psi^\mathrm{I}(2)
   &=E\psi^\mathrm{I}(1) \\
   t_0\psi^\mathrm{I}(1)+(t_2+\gamma)\psi^\mathrm{II}(2)
   &=E\psi^\mathrm{II}(1)
\end{aligned}
\right.
\text{\quad and \quad}
\left\{
\begin{aligned}
   t_0\psi^\mathrm{II}(N)+(t_1+\gamma)\psi^\mathrm{I}(N-1)
   &=E\psi^\mathrm{I}(N) \\
   t_0\psi^\mathrm{I}(N)+(t_2-\gamma)\psi^{\mathrm{II}}(N-1)
   &=E\psi^\mathrm{II}(N)
\end{aligned}.
\right.
\label{eq:GBZbc}
\ee
Replacing $E$ using the bulk relations in \cref{eq:abroots} leads to the following four linear equations:
\be
\left\{
\begin{aligned}
   -(t_1+\gamma)c_1-(t_1+\gamma)c_2
   +t_0\beta_1 d_1+t_0\beta_2 d_2=0 \\
   t_0\alpha_1 c_1+t_0\alpha_2c_2
   -(t_2-\gamma)d_1-(t_2-\gamma)d_2=0\\
   -(t_1-\gamma)\alpha_1^{N+1}c_1
   -(t_1-\gamma)\alpha_2^{N+1}c_2
   +t_0\beta_1^Nd_1+t_0\beta_2^Nd_2=0\\
   t_0\alpha_1^Nc_1+t_0\alpha_2^Nc_2
   -(t_2+\gamma)\beta_1^{N+1}d_1
   -(t_2+\gamma)\beta_2^{N+1}d_2=0.
\end{aligned}
\right.
\ee
We can express these equations in matrix form as $M(c_1,c_2,d_1,d_2)^T=0$, where a nontrivial solution exists only when the determinant vanishes:
\be
\text{det}(M)=
\begin{vmatrix}
-(t_1+\gamma) & -(t_1+\gamma)
& t_0\beta_1 & t_0\beta_2\\
t_0\alpha_1 & t_0\alpha_2
& -(t_2-\gamma) & -(t_2-\gamma)\\
-(t_1-\gamma)\alpha_1^{N+1}
& -(t_1-\gamma)\alpha_2^{N+1}
& t_0\beta_1^N & t_0\beta_2^N\\
t_0\alpha_1^N & t_0\alpha_2^N
& -(t_2+\gamma)\beta_1^{N+1}
& -(t_2+\gamma)\beta_2^{N+1}
\end{vmatrix}=0.
\ee
For the parameters considered here, $t_1+\gamma\ne0$, and using Schur's determinant identity,
$\text{det}\begin{pmatrix}A&B\\C&D\end{pmatrix}
=\text{det}(A)\text{det}(D-CA^{-1}B)$,
and choosing $A=-(t_1+\gamma)$, we obtain
\be
\text{det}(D-CA^{-1}B)=
\text{det}\begin{pmatrix}
t_0(\alpha_2-\alpha_1)
& -(t_2-\gamma)+\frac{t_0^2}{t_1+\gamma}\alpha_1\beta_1
& -(t_2-\gamma)+\frac{t_0^2}{t_1+\gamma}\alpha_1\beta_2\\
(t_1-\gamma)(\alpha_1^{N+1}-\alpha_2^{N+1})
& t_0\beta_1^N-\frac{t_0(t_1-\gamma)}{t_1+\gamma}\alpha_1^{N+1}\beta_1
& t_0\beta_2^N-\frac{t_0(t_1-\gamma)}{t_1+\gamma}\alpha_1^{N+1}\beta_2\\
t_0(\alpha_2^N-\alpha_1^N)
& -(t_2+\gamma)\beta_1^{N+1}+\frac{t_0^2\alpha_1^N\beta_1}{t_1+\gamma}
& -(t_2+\gamma)\beta_2^{N+1}+\frac{t_0^2\alpha_1^N\beta_2}{t_1+\gamma}
\end{pmatrix}=0.
\ee
Since we consider weak coupling $t_0$, the $t_0^2$ terms in the first row can be neglected compared with $t_2-\gamma$
provided $t_0^2|\alpha_1\beta_j|\ll(t_1+\gamma)(t_2-\gamma)$ for $j=1,2$.
Moreover,
we restrict attention to the large-$N$ regime where $|\alpha_1/\alpha_2|^N\ll1$ and $|\beta_1/\beta_2|^N\ll1$.
With these approximations, the determinant simplifies to
\be
\det(D-CA^{-1}B)\approx
\text{det}\begin{pmatrix}
t_0(\alpha_2-\alpha_1)
& -(t_2-\gamma) & -(t_2-\gamma)\\
-(t_1-\gamma)\alpha_2^{N+1}
& t_0\beta_1^N-\frac{t_0(t_1-\gamma)}{t_1+\gamma}\alpha_1^{N+1}\beta_1
& t_0\beta_2^N-\frac{t_0(t_1-\gamma)}{t_1+\gamma}\alpha_1^{N+1}\beta_2\\
t_0\alpha_2^N
& -(t_2+\gamma)\beta_1^{N+1}+\frac{t_0^2\alpha_1^N\beta_1}{t_1+\gamma}
& -(t_2+\gamma)\beta_2^{N+1}+\frac{t_0^2\alpha_1^N\beta_2}{t_1+\gamma}
\end{pmatrix}=0.
\label{eq:S11}
\ee
Expanding Eq.~\eqref{eq:S11} gives terms proportional to
$\alpha_1^N\alpha_2^N$, $\alpha_1^N\beta_1^N$,
$\alpha_1^N\beta_2^N$, $\alpha_2^N\beta_1^N$,
$\alpha_2^N\beta_2^N$, and $\beta_1^N\beta_2^N$.
Taking into account the ordering of $|\alpha_{1,2}|$ and $|\beta_{1,2}|$,
together with the relation in Eq.~\eqref{eq:abroots}, the dominant terms are
$\alpha_1^N\alpha_2^N$ and $\alpha_2^N\beta_2^N$.
Keeping only these two dominant terms and canceling the common factor
$(t_2-\gamma)\alpha_2^N$, we obtain
\bea
\frac{t_0^2(t_1-\gamma)}{t_1+\gamma}
(\alpha_1-\alpha_2)(\beta_2-\beta_1)\alpha_1^N
+
\left[
-t_0^2+(t_1-\gamma)(t_2+\gamma)\alpha_2\beta_2
\right]\beta_2^N
\simeq 0.
\label{eq:S12}
\eea
Recalling the definition of the GBZ \cite{yao2018edge,yokomizo2019non},
we observe that this equation contains only two surviving exponential factors,
$\alpha_1^N$ and $\beta_2^N$.
Therefore, within the present approximation, the effective finite-size
GBZs for the two chains are determined by the corresponding roots
$\alpha_1$ and $\beta_2$, respectively.
Furthermore, in the weak coupling regime,
$t_0^2\ll\left|(t_1-\gamma)(t_2+\gamma)\alpha_2\beta_2\right|$.
Neglecting the $t_0^2$ term, we obtain
\bea
\left(\frac{\beta_2}{\alpha_1}\right)^N
\simeq
t_0^2
\frac{(\alpha_2-\alpha_1)(\beta_2-\beta_1)}
{(t_1+\gamma)(t_2+\gamma)\alpha_2\beta_2},
\qquad
\text{i.e.}\quad
\left|\frac{\beta_2}{\alpha_1}\right|
\simeq
\left[
\frac{t_0^2}
{(t_1+\gamma)(t_2+\gamma)}
\right]^{1/N}.
\label{eq:GBZbeta}
\eea
where the $O(1/N)$ correction to $\ln|\beta_2/\alpha_1|$,
$\frac{1}{N}
\ln\left|\frac{(\alpha_2-\alpha_1)(\beta_2-\beta_1)}
{\alpha_2\beta_2}\right|$,
has been omitted
in the second relation, assuming that this logarithm remains $O(1)$.

Combining Eq.~\eqref{eq:GBZbeta} with Eq.~\eqref{eq:abroots},
we can solve for the root moduli $|\alpha_1|$ and $|\beta_2|$.
In particular, for the symmetric case $t_1=t_2$ used in the main article,
the relationship simplifies to $\beta_2=\alpha_1^{-1}$.
For this symmetric case, Eq.~\eqref{eq:GBZbeta} therefore gives the following GBZs:
\be
\left\{
\begin{aligned}
\alpha&\approx
\left[\frac{t_0}{(t_1+\gamma)}\right]^{-1/N}
e^{-i\theta},
\quad \text{for chain I}\\
\beta&\approx
\left[\frac{t_0}{(t_1+\gamma)}\right]^{1/N}
e^{i\theta},
\quad \text{for chain II},
\end{aligned}
\right.
\qquad
\text{with }\text{continuous parameter }
\theta\in[0,2\pi].
\ee
By substituting these GBZ expressions into the bulk equations, we obtain the approximate analytical spectrum
\be
E_{\rm ana}
=(t_2+\gamma)
\left[\frac{t_0}{(t_1+\gamma)}\right]^{1/N}e^{i\theta}
+(t_2-\gamma)
\left[\frac{t_0}{(t_1+\gamma)}\right]^{-1/N}e^{-i\theta}.
\label{eq:E_ana}
\ee
Writing $r=[t_0/(t_1+\gamma)]^{1/N}=e^{-\frac{1}{N}\ln\frac{t_1+\gamma}{t_0}}$,
Eq.~\eqref{eq:E_ana} gives
$\max_{\theta}\operatorname{Im}E_{\rm ana}
=|(t_2+\gamma)r-(t_2-\gamma)r^{-1}|$.
In the large-$N$ regime, expanding $r$ to first order in $1/N$ gives
$\max_{\theta}\operatorname{Im}E_{\rm ana}=2\left[\gamma-\frac{t_2}{N}\ln\frac{t_1+\gamma}{t_0}\right]
+O(N^{-2})$.

As shown in \cref{fig:supp-verification-GBZ},
the eigenvalue spectra for the active two-particle sector
in Eq.~\eqref{eq:supp-model-gbz} agree well with $E_{\rm ana}$
for the parameter choices shown.

\section{SIII. Estimation of scaling behaviour of four-particle system}

In this section, we provide a simple but intuitive estimate for the finite-size scaling of \(\max \operatorname{Im}E\) in the four-particle sector [Fig.~1(d1) of the main text].
For the four-particle system ($n=4$), bosonic indistinguishability connects
cNHSE-like branches into global feedback loops in the Hilbert space
connectivity graph of the OBC Hamiltonian [Fig.~1(b3) of the main text].
The full graph contains many coupled loops and cannot be reduced exactly to a
single two-chain ladder. To obtain a scaling estimate, we isolate a representative feedback path (e.g., the largest loop in Fig.~1(b3) of the main text) comprising $L$ ordered Fock configurations ($\lvert \mathcal F_j\rangle$). Here, $j$ labels position along the effective
path rather than a real-space site. The path is thus described by
\begin{align}
 H_{\mathrm{loop}}={}&\sum_{j=1}^{L-1}\left(
 t_+^{\mathrm{eff}}\lvert\mathcal F_{j+1}\rangle
 \langle\mathcal F_j\rvert
 +t_-^{\mathrm{eff}}\lvert\mathcal F_j\rangle
 \langle\mathcal F_{j+1}\rvert\right) \notag\\
 &+t_+^{\mathrm b}\lvert\mathcal F_1\rangle
 \langle\mathcal F_L\rvert
 +t_-^{\mathrm b}\lvert\mathcal F_L\rangle
 \langle\mathcal F_1\rvert .
 \label{eq:Hloop}
\end{align}
Here, $t_\pm^{\mathrm{eff}}>0$ denote the effective hopping amplitudes along the path in the Hilbert-space connectivity graph, whereas $t_\pm^{\mathrm b}>0$ close the loop. They need not
equal the microscopic hoppings $t_1\pm\gamma$ [\cref{eq:supp-model}] because of particle statistics.

The clockwise and counterclockwise hopping products [Fig.~1(b3) of the main text] are
\begin{equation}
 P_+=\left(t_+^{\mathrm{eff}}\right)^{L-1}t_+^{\mathrm b},
 \qquad
 P_-=\left(t_-^{\mathrm{eff}}\right)^{L-1}t_-^{\mathrm b}.
 \label{eq:directional_products}
\end{equation}
With Fourier transformation,
$\lvert k_m\rangle=L^{-1/2}\sum_j e^{ik_mj}\lvert\mathcal F_j\rangle$, one
obtains the spectra
\begin{equation}
 E^{\mathrm{loop}}
 =\bar t_{+,L}e^{-ik_m}
 +\bar t_{-,L}e^{ik_m},
 \qquad
 k_m=\frac{2\pi m}{L},
 \quad m=0,1,\ldots,L-1,
 \label{eq:loop_spectrum}
\end{equation}
where we approximate the inhomogeneous loop by a homogeneous ring with effective directional hoppings $\bar t_{\pm,L}=P_\pm^{1/L}
 =t_\pm^{\mathrm{eff}}
 \left(\frac{t_\pm^{\mathrm b}}
 {t_\pm^{\mathrm{eff}}}\right)^{1/L}$.
For $\bar t_{+,L}>\bar t_{-,L}$, its largest imaginary part is
\begin{equation}
 \max\operatorname{Im}E^{\mathrm{loop}}
 =
 \left(\bar t_{+,L}-\bar t_{-,L}\right)q_L,
 \qquad
 q_L\equiv\max_m|\sin k_m|,
 \label{eq:max_imag_loop}
\end{equation}
where $q_L=1+O(L^{-2})$. 
For larger $L$, we have
$\bar t_{\pm,L}
 =t_\pm^{\mathrm{eff}}
 +\frac{t_\pm^{\mathrm{eff}}}{L}
 \ln\left(\frac{t_\pm^{\mathrm b}}
 {t_\pm^{\mathrm{eff}}}\right)
 +O(L^{-2}),$
and hence
\begin{equation}
 \max\operatorname{Im}E^{\mathrm{loop}}
 =t_+^{\mathrm{eff}}-t_-^{\mathrm{eff}}
 +\frac{1}{L}
 \left[
 t_+^{\mathrm{eff}}
 \ln\left(\frac{t_+^{\mathrm b}}{t_+^{\mathrm{eff}}}\right)
 -
 t_-^{\mathrm{eff}}
 \ln\left(\frac{t_-^{\mathrm b}}{t_-^{\mathrm{eff}}}\right)
 \right]
 +O(L^{-2}).
 \label{eq:L_scaling}
\end{equation}

Suppose that a representative global path traverses a fixed number of
$O(N)$-long branches, so that $L=\alpha N+O(1)$, and that the corresponding
loop controls the largest imaginary part of eigenvalues. \cref{eq:L_scaling} then predicts
\begin{equation}
 \max\operatorname{Im}E(N)
\simeq
 M_\infty-\frac{A}{N}+O(N^{-2}),
 \label{eq:N_scaling}
\end{equation}
where $N$ is the system size, $M_\infty$ and $A$ can be fitted.

\begin{figure}[htbp]
 \centering
 \includegraphics[width=0.98\textwidth]{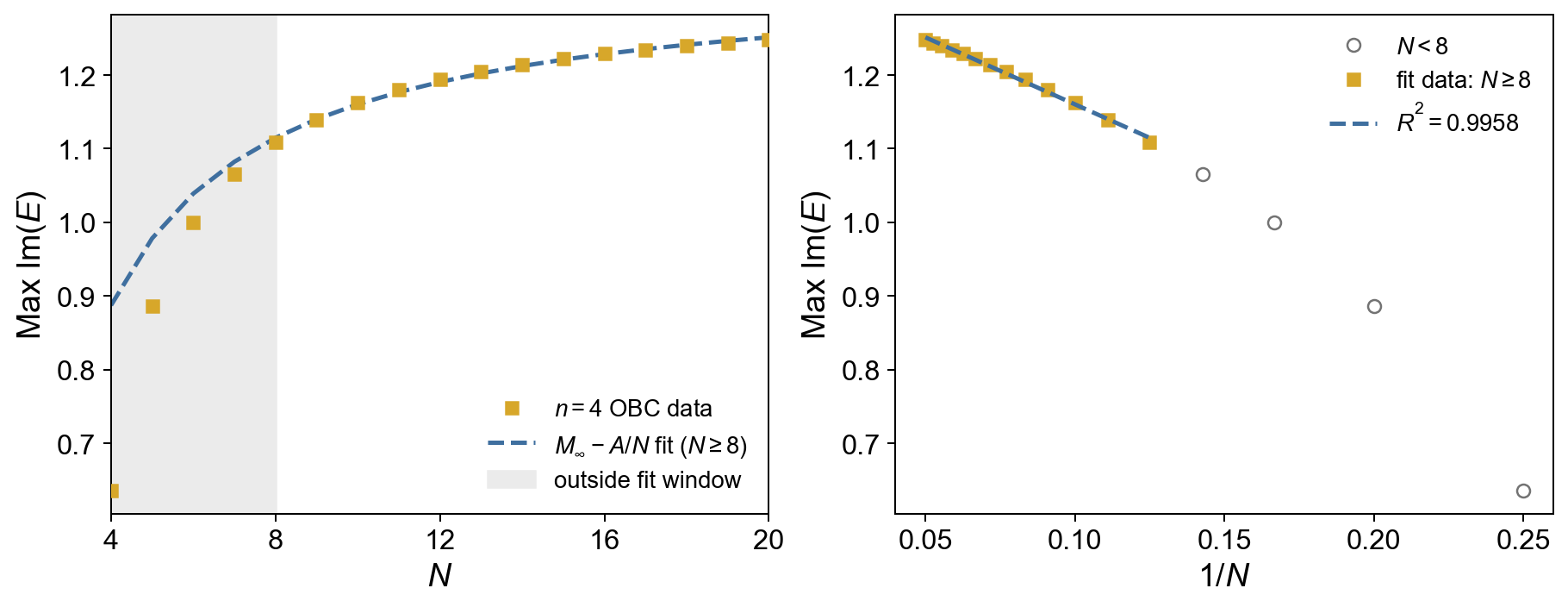}
 \caption{Finite-size dependence of $\max\operatorname{Im}(E)$ for the
 four-particle system ($n=4$) [\cref{eq:supp-model}] under OBCs. Left: exact-diagonalization data
 from Fig.~1(d1) and a fit to $M_\infty-A/N$ [\cref{eq:N_scaling}] over $N\geq8$. The gray region
 is excluded from the fit, and the dashed curve within it is an
 extrapolation.  Right: the same data plotted against
 $1/N$. The finite-window fit gives $M_\infty=1.3421$, $A=1.8181$, and
 $R^2=0.9958$. Parameters are $t_0=10^{-2}$, $t_1=0.58$, $\gamma=0.25$,
 and $n_{\mathrm{occ}}^{\max}=2$.}
 \label{fig:n4_scaling}
\end{figure}

As shown in \cref{fig:n4_scaling}, the exact-diagonalization results for the
four-particle system are well described by the finite-size form in
\cref{eq:N_scaling} for $N\geq8$. In particular,
$\max\operatorname{Im}E$ increases with $N$ and approaches a finite
asymptotic value, while plotting the same data against $1/N$ yields an
approximately linear dependence. A fit to $M_\infty-A/N$ gives
$M_\infty=1.3421$, $A=1.8181$, and $R^2=0.9958$. The deviations at smaller
$N$ are consistent with the expected finite-size corrections beyond the
leading $1/N$ term. Thus, although the full four-particle Hilbert-space graph
contains many coupled feedback loops, the representative-loop construction
captures the observed leading scaling of $\max Im (E)$

\newpage

\section{SIV. Robustness and generality of scaling-induced real-to-complex transition}

In this section, we illustrate that scaling-induced real-to-complex transition is not a fine-tuned feature of the particular model parameters considered in the main text. We first identify the inter-clustering coupling $t_0$ as an essential ingredient of the spectral complexification mechanism. We then demonstrate that the phenomenon persists upon:
\begin{itemize}
\item removing the occupancy constraint
\item varying the hopping parameters and diagnostic cutoffs
\item increasing the correlated-hopping range
\item extending the interaction from two-body to three-body processes. 
\end{itemize}
These results show that the real-to-complex transition represents a robust and broadly applicable many-body mechanism rather than a special property of a single parameter choice or microscopic Hamiltonian.

\begin{figure}[htbp]
    \centering
    \includegraphics[width=0.9\linewidth]{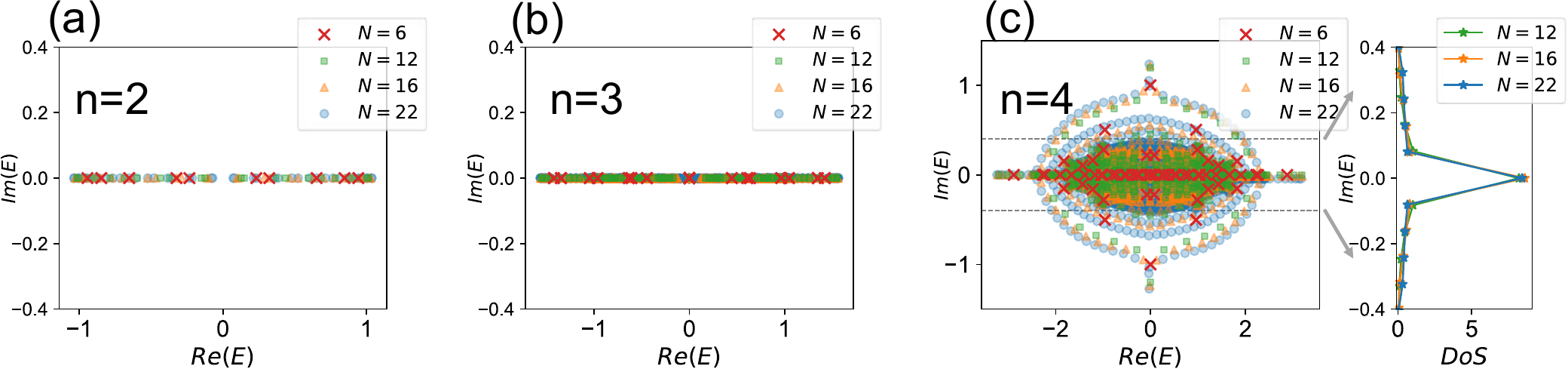}
    \caption{Eigenspectra of our model [\cref{eq:supp-model}] for $t_0=0$ under OBCs, shown for various system sizes $N$, with a maximum on-site occupancy $n^{max}_{occ}=2$ for two-($n=2$), three-($n=3$) and four-($n=4$) particle systems.
    The parameters are fixed at $t_1 = 0.58$, and $\gamma = 0.25$.}
    \label{fig:supp-spectra-t0_0}
\end{figure}

\noindent\textbf{Role of the symmetric coupling $t_0$:}
Firstly, we illustrate the crucial role of the correlated hopping term $t_0$ [\cref{eq:supp-model}] in driving the real-to-complex transition. In \cref{fig:supp-spectra-t0_0}, we show that for $t_0 = 0$, the spectrum remains entirely real for particle numbers $n = 2$ and $n = 3$, indicating the complete absence of real-to-complex transition. For the case with $n = 4$ particles, where the spectrum remains complex throughout, we examine the system using the density of states (DOS) resolved along the imaginary axis $\operatorname{Im}(E)$. As shown in \cref{fig:supp-spectra-t0_0}(c), the DOS does not broaden with increasing system size $N$, indicating the absence of competitive skin mechanism -- the complex spectrum in the $n=4$ case is simply due to the presence of global loops in the Hilbert space graph [see Fig.~1 of the main text].  Taken together, these results confirm that the real-to-complex transition and the associated scaling behavior from skin competition within the Hilbert space ladders require a nonzero $t_0$.

\begin{figure}[htbp]
    \centering
    \includegraphics[width=\linewidth]{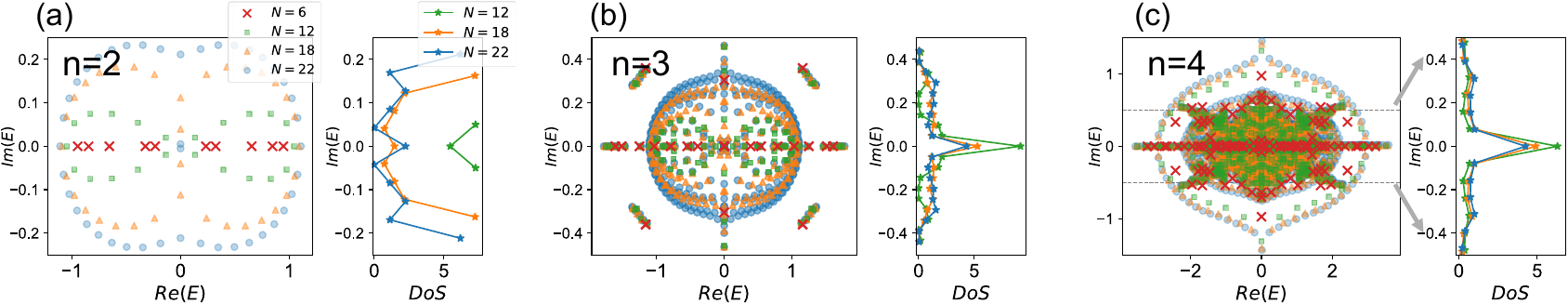}
    \caption{With no occupancy limit, spectra complexification of our model [\cref{eq:supp-model}] with system sizes $N$ persists for two- ($n=2$), three- ($n=3$) and four- ($n=4$) particle systems under OBCs.
    The parameters used are $t_0 = 10^{-2}$, $t_1 = 0.58$, and $\gamma = 0.25$.}
    \label{fig:supp-spectra-nolimit}
\end{figure}

\noindent\textbf{Robustness against the occupancy constraint:}
Having established the essential role of $t_0$, we next show that the real-to-complex transition does not depend on the imposed occupancy constraint. As shown in \cref{fig:supp-spectra-nolimit}, the $\operatorname{Im}(E)$-resolved DOS broadens with increasing system size $N$ for two-, three-, and four-particle systems even when the occupancy constraint is removed. The persistence of this scaling behavior demonstrates that the real-to-complex transition originates from the underlying correlated-hopping structure rather than a feature specific to a particular choice of local constraint.

\begin{figure}[H]
    \centering
    \includegraphics[width=\linewidth]{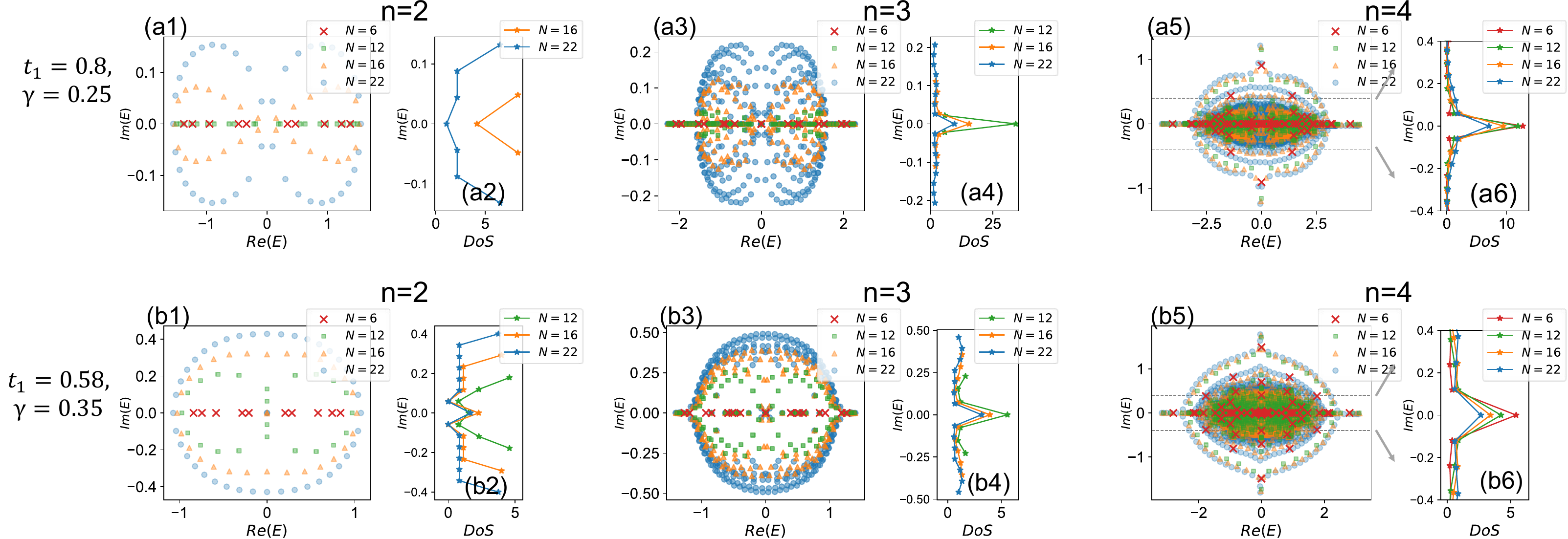}
    \caption{Additional eigenspectra and density of states (DOS) resolved by the imaginary parts of the eigenvalues, $\operatorname{Im}(E)$, for two-, three-, and four-particle systems ($n=2,3,4$) at various system sizes $N$, demonstrating that the real-to-complex transition is generic for different parameter choices. We fix $t_0=10^{-2}$ and impose a maximum on-site occupancy $n_{\mathrm{occ}}^{\max}=2$, with OBCs throughout. In \textbf{(a)}, $t_1=0.8$ and $\gamma=0.25$; in \textbf{(b)}, $t_1=0.58$ and $\gamma=0.35$.}
    \label{fig:supp-spectra-t1-gamma}
\end{figure}

\begin{figure}[H]
    \centering
    \includegraphics[width=0.9\linewidth]{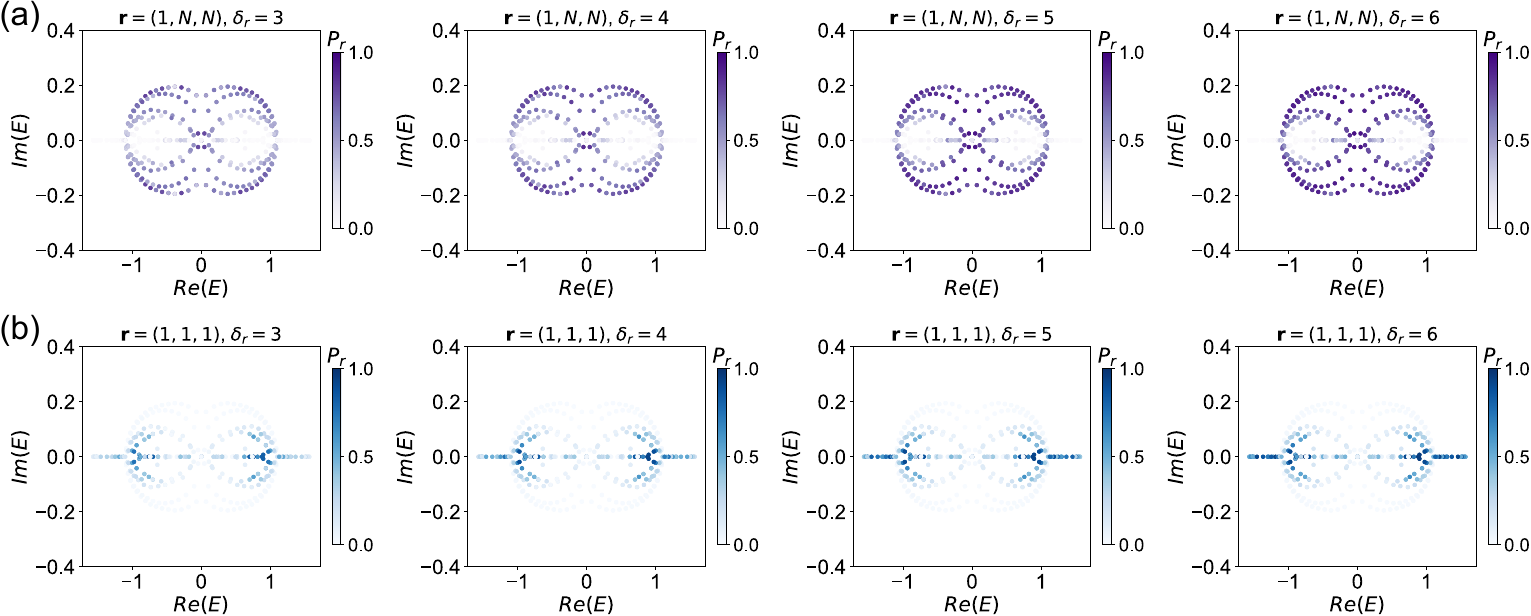}
    \caption{Supplementary spectra corresponding to Fig.~2(a1-a2) in the main text for different $\delta_{\bm r}$. The many-body unipolar and asymmetric bipolar localizations are robust against variations in $\delta_{\bm r}$ [\cref{eq:supp-local-prob}]. All other parameters are the same as in Fig.~2 of the main text: $t_1 = 0.58$, $\gamma = 0.25$, and $t_0 = 10^{-3}$, with particle number $n=3$ and system size $N=20$, OBCs imposed throughout, and maximum on-site occupation $n_{\mathrm{occ}}^{\max}=2$. }
    \label{fig:supp-prob-r}
\end{figure}

\noindent\textbf{Robustness against parameter variations:}
In the main text, we used a representative parameter set to demonstrate the onset of \emph{real-to-complex transition}, i.e., the transition from a predominantly real to a broadly complex energy spectrum.
Here, we provide complementary results showing that this phenomenon persists over a broad range of parameter choices:
\begin{itemize}[noitemsep, topsep=0pt]
    \item it persists for other choices of $(t_1,\gamma)$ and for different particle numbers $n$ (shown in \cref{fig:supp-spectra-t1-gamma}, complementary to Fig.~1(c) in the main text);
    \item the probability-resolved spectra employed to diagnose localization patterns remain stable upon varying the coarse-graining cutoff $\delta_{\bm r}$ entering their definition [shown in \cref{fig:supp-prob-r}, complementary to Fig.~2(a1-a2) in the main text].
\end{itemize}

\cref{fig:supp-spectra-t1-gamma} shows eigenvalue spectra in the complex plane (left panels) together with the density of states (DOS) resolved by the imaginary part of the eigenvalues (right panels), for $n=2,3,4$ bosons and several system sizes $N$ (symbols in the legends).

For each $n$, the spectra evolve from being mostly real at small $N$ to exhibiting pronounced complex structure at larger $N$.
Concretely, for $n=2$ [panels (a1),(b1)] the eigenvalues progressively move off the real axis and form symmetric arcs/loops in $\mathrm{Im}(E)$, while the corresponding $\mathrm{Im}(E)$-resolved DOS [panels (a2),(b2)] develops clear weight at finite $|\mathrm{Im}(E)|$ rather than being concentrated at $\mathrm{Im}(E)=0$.
For $n=3$ [panels (a3),(b3)], the complex loops become broader and more densely populated as $N$ increases, and the DOS profiles [panels (a4),(b4)] show enhanced support away from $\mathrm{Im}(E)=0$.
For $n=4$ [panels (a5),(b5)], the complex spectrum further expands into a two-dimensional region in the $(\mathrm{Re}\,E,\mathrm{Im}\,E)$ plane, with a correspondingly broad DOS in $\mathrm{Im}(E)$ [panels (a6),(b6)].
Importantly, this behavior is observed for both parameter sets (rows a and b), demonstrating that a real-to-complex transition is not tied to a fine-tuned choice of $(t_1,\gamma)$ and is consistently present across particle numbers.

To quantify such unipolar/bipolar localization of an eigenstate $\psi$, we define the eigenstate weight within a neighborhood of a reference configuration $\bm r$ as
\begin{equation}
    P_{\bm r}(\psi) = \sum_{(x_1,x_2,x_3) \in \mathcal{C}_{\bm r}} |\psi(x_1,x_2,x_3)|^2,
    \label{eq:supp-local-prob}
\end{equation}
where $\mathcal{C}_{\bm r}=\{(x_1,x_2,x_3)\,|\,\|(x_1,x_2,x_3)-\bm r\|_2\le \delta_{\bm r}\}$ is the set of configurations whose Euclidean distance (in lattice units) from $\bm r$ does not exceed $\delta_{\bm r}$. Here $\bm r\in\{(1,1,1),(1,1,N),(1,N,N),(N,N,N)\}$ labels representative edge configurations. Thus, $P_{\bm r}$ provides a quantitative measure of the degree of unipolar/bipolar localization exhibited by a given eigenstate.

\Cref{fig:supp-prob-r} shows spectra colored by $P_{\bm r}$ for several choices of the coarse-graining scale $\delta_{\bm r}$, demonstrating that the resulting localization diagnosis is robust. For all $\delta_{\bm r}$ (and across the parameter sets considered), the eigenstates can be systematically grouped into three regimes, consistent with the main-text discussion: (i) \emph{asymmetric bipolar} states localized predominantly near $(1,N,N)$ [purple region in \cref{fig:supp-prob-r}(a)]; (ii) \emph{unipolar} states localized near $(1,1,1)$ [blue region in \cref{fig:supp-prob-r}(b)]; and (iii) a \emph{mixed} regime comprising hybrid states with substantial weight near both configurations [overlap of \cref{fig:supp-prob-r}(a) and \cref{fig:supp-prob-r}(b)].

\noindent\textbf{Generalizations to longer-range correlated hopping and higher-body interactions:}
Finally, we test whether the real-to-complex transition extends beyond the microscopic Hamiltonian considered in the main text. We generalize the correlated-hopping distance to an arbitrary value $d$ and subsequently extend the construction to three-body correlated interactions. These generalizations are important because they determine whether the observed complexification is specific to the original two-body, short-range model or reflects a more general organizing principle for interacting non-Hermitian systems.

\begin{figure}[htbp]
    \centering
    \includegraphics[width=\linewidth]{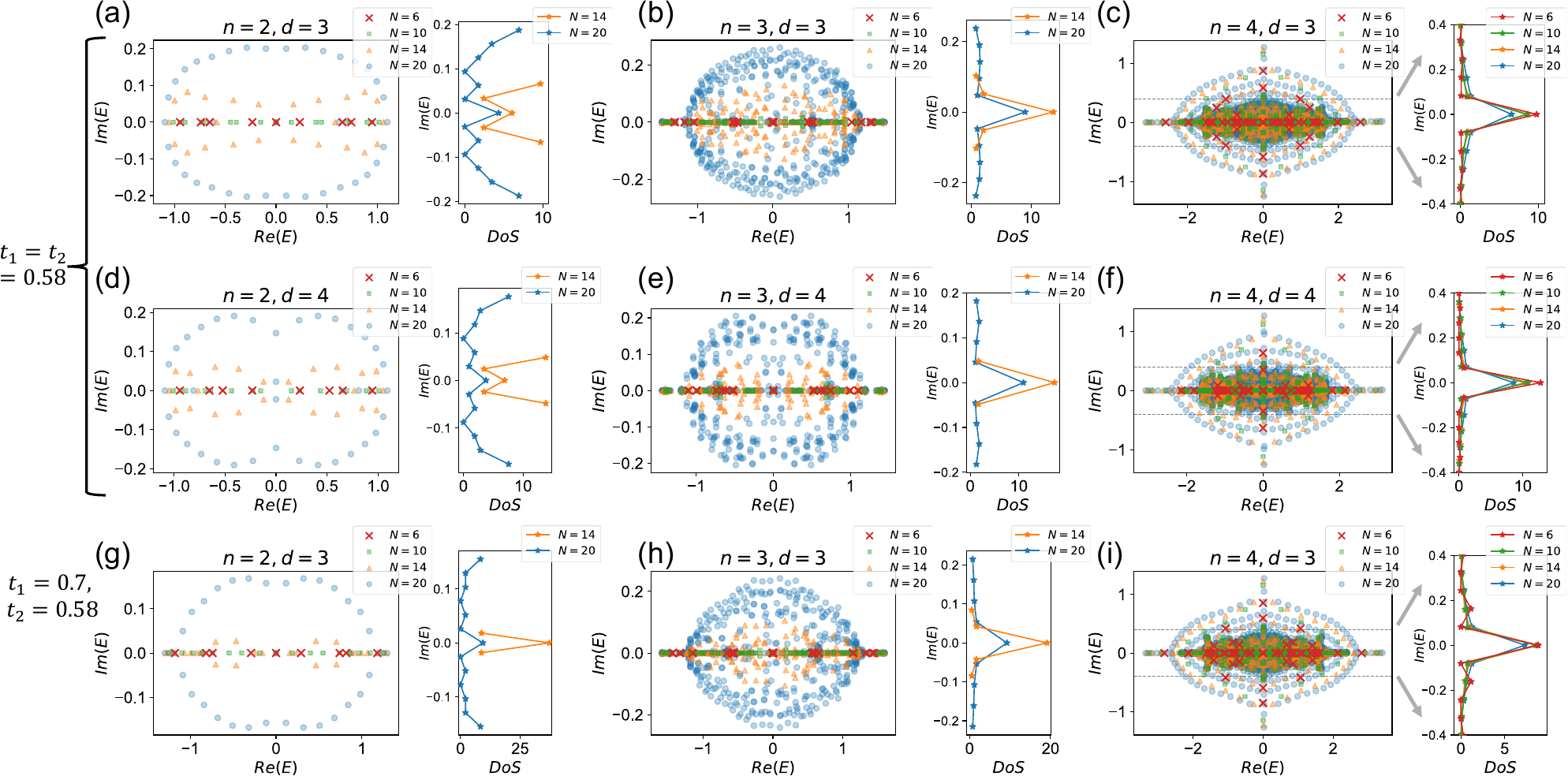}
    \caption{Real-to-complex transition in the generalized model \cref{eq:supp-gen-model-1} for $d=3,4$, shown for two-, three-, and four-particle systems ($n=2,3,4$) at various system sizes $N$ under the occupancy constraint $n_{\mathrm{occ}}^{\max}=2$.
    Panels \tb{(a)--(f)} use hopping parameters $t_1=t_2=0.58$, whereas panels \tb{(g)--(i)} use $t_1=0.7$ and $t_2=0.58$.
    Throughout, we fix $t_0=10^{-2}$ and $\gamma=0.25$ and impose OBCs. }
    \label{fig:supp-gen-model-1}
\end{figure}

The generalized Hamiltonian with correlated hopping distance $d$ is given by
\begin{equation}
\begin{aligned}
H &= \sum_{i=1}^{N} \frac{t_0}{\sqrt{2}} c_{i+1}^{\dagger}c_{i}c_{i+1-d}^{\dagger}c_{i} + \frac{t_{0}}{\sqrt{2}}c_{i}^{\dagger}c_{i+1-d}c_i^{\dagger}c_{i+1} +\frac{t_{1} + \gamma}{2}c_{i+1}^{\dagger}c_{i}c_{i+1}^{\dagger}c_{i} + \frac{t_{1} - \gamma}{2}c_{i}^{\dagger}c_{i+1}c_{i}^{\dagger}c_{i+1} \\
&+(t_{2} - \gamma)c_{i+2}^{\dagger}c_{i+1}c_{i+2-d}^{\dagger}c_{i+1-d} + (t_{2} + \gamma)c_{i+1-d}^{\dagger}c_{i+2-d}c_{i+1}^{\dagger}c_{i+2}.
\end{aligned}
\label{eq:supp-gen-model-1}
\end{equation}
As shown in \cref{fig:supp-gen-model-1}, a real-to-complex transition persists across a range of hopping distances $d$ and particle numbers $n$. In particular, the eigenvalue distribution exhibits significant broadening along the imaginary axis. This is quantitatively confirmed by the density of states (DOS) resolved by the imaginary part of the eigenvalues, which shows a clear spread along the $\operatorname{Im}(E)$ axis—indicating the robustness of the real-to-complex transition. This universality exists because changing $d$ only changes the real-space diameters of the clusters, and not the qualitative structure of the Hilbert space.

\begin{figure}[htbp]
    \centering
    \includegraphics[width=\linewidth]{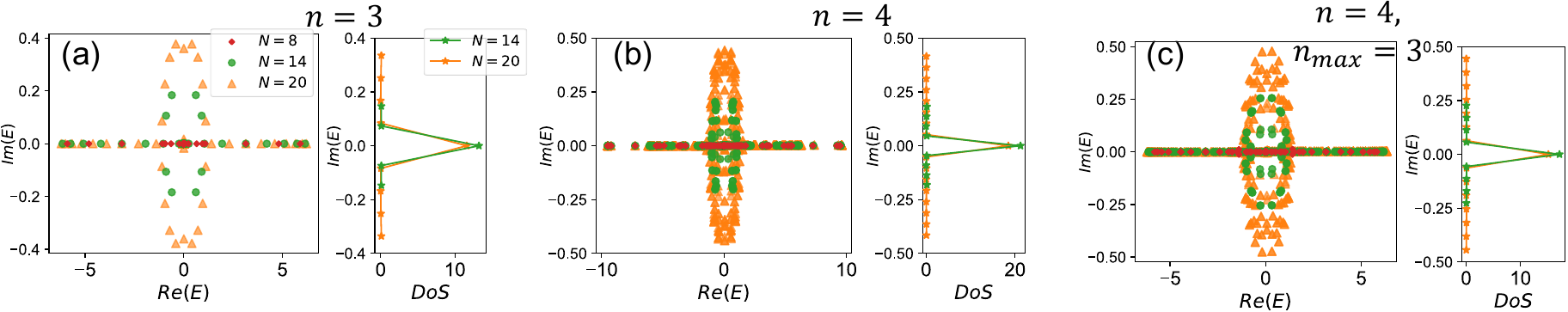}
    \caption{Real-to-complex transition in the generalized three-body interaction model [\cref{eq:supp-gen-model-2}] for \tb{(a)} $n=3$ without an occupancy constraint,
    \tb{(b)} $n=4$ without an occupancy constraint, and
    \tb{(c)} $n=4$ with the occupancy constraint $n_{\mathrm{occ}}^{\max}=3$,
    shown for various system sizes $N$.
    Parameters are $t_0=10^{-2}$, $t_1=t_2=0.58$, and $\gamma=0.25$, and we impose OBCs throughout.}
    \label{fig:supp-gen-model-2}
\end{figure}

Furthermore, we can also extend our study to a three-body interaction, for which a generalized Hamiltonian is given by 
\begin{equation}
\begin{aligned}
    H_{3-body} &= \sum_{i=1}^{N} (t_1+\gamma) c_{i+1}^{\dagger}c_i c_{i+1}^{\dagger}c_i c_{i+1}^{\dagger}c_i + (t_1-\gamma)c_{i}^{\dagger}c_{i+1} c_{i}^{\dagger}c_{i+1} c_{i}^{\dagger}c_{i+1} \\
    & + (t_2-\gamma) c_{i}^{\dagger}c_{i-1} c_{i+1}^{\dagger}c_i c_{i+2}^{\dagger}c_{i+1} + (t_2+\gamma)c_{i+1}^{\dagger}c_{i+2} c_{i}^{\dagger}c_{i+1} c_{i-1}^{\dagger}c_{i} \\
    &+ t_0 c_{i-1}^{\dagger}c_{i} c_{i}^{\dagger}c_i c_{i+1}^{\dagger}c_{i} + t_0c_{i}^{\dagger}c_{i+1} c_{i}^{\dagger}c_{i} c_{i}^{\dagger}c_{i-1}.
    \label{eq:supp-gen-model-2}
\end{aligned}
\end{equation}
Similarly, \cref{fig:supp-gen-model-2} presents the spectra of the system governed by \cref{eq:supp-gen-model-2} for (a) $n=3$ particles, (b) $n=4$ particles without any occupancy constraints, and  (c) $n=4$ particles with a maximum on-site occupancy $n_{\mathrm{occ}}^{\max}=3$. In all three cases,
the spectrum progressively broadens along $\operatorname{Im}(E)$ as the system size $N$ increases.

For $n=3$, the three-body processes generate the oppositely biased cluster-translation channels $(i,i,i)\rightleftharpoons(i+1,i+1,i+1)$ and $(i-1,i,i+1)\rightleftharpoons(i,i+1,i+2)$, which are coupled through the $t_0$ processes. Together, they form a cNHSE-like coupled-chain motif at the level of three-particle clusters.
For $n=4$, a global ladder mapping no longer applies, but these channels remain local building blocks whose $t_0$-mediated feedback loops produce the same size-dependent broadening. 

Taken together, the results in this section show that scaling-induced real-to-complex transition is insensitive to the occupancy constraint, persists across different hopping parameters and diagnostic cutoffs, and survives substantial modifications of the microscopic dynamics, including longer-range hopping and higher-body interactions. At the same time, the disappearance of the scaling behavior when $t_0=0$ identifies the coupling between distinct hopping patterns as the essential structural ingredient. A real-to-complex transition should therefore be understood not as a fine-tuned feature of the model studied in the main text, but as a generic many-body phenomenon arising from coupled nonreciprocal hopping structures.

\section{SV. Eigenstate signatures of real-to-complex transition from entanglement entropy and inverse participation
ratio}
Beyond the eigenvalue spectra, this section examines how the real-to-complex transition is reflected in the structure of the corresponding many-body eigenstates.
As $t_0$ increases, eigenvalues become more ''complex''. Simultaneously, the eigenstates become less localized in Hilbert space due to the antagonism between competing NHSE channels: their inverse participation ratio (IPR) decreases and their bipartite entanglement entropy $S_{\mathrm{EE}}$ increases. Together, these two diagnostics provide a consistent picture of a crossover from weakly-entangled, Hilbert space-localized eigenstates to strongly-entangled, delocalized eigenstates.

\subsection{Entanglement Entropy $S_{EE}$ in a many-body bosonic system }

We consider a one-dimensional bosonic lattice system with $N$ sites and a total of $n$ bosons. The many-body Hilbert space is spanned by Fock states of the form $\ket{n_1, n_2,...,n_{i},..., n_{N-1}, n_N}$, where $n_i$ denotes the occupation number at site $i$, subject to the constraint $\sum_{i=1}^N n_i= n$. To calculate the entanglement entropy, we bipartition the system into two subsystems, $A$ and $B$, with subsystem $A$ consisting of the first $L=\lfloor N/2\rfloor$ sites and subsystem $B$ the remaining $N-L$ sites. We denote the occupation numbers in $A$ as $\{n_1, n_2, \dots, n_L\}$ and in $B$ as $\{n_{L+1}, \dots, n_N\}$.

For each (right) eigenstate $\ket{\psi}$ associated with $E$, $H\ket{\psi}=E \ket{\psi}$,  we can write the pure-state density matrix as
\begin{equation}
\rho=\ket{\psi}\bra{\psi},
\end{equation}
where $\langle\psi|\psi\rangle = 1$ is assumed.
We emphasize that the entanglement entropy throughout the work is computed using only the right eigenbasis, in contrast to that defined in a bi-orthogonal basis~\cite{kawabata2023entanglement,lee2024entanglement,lee2022exceptional,chang2020entanglement}.
Then we trace out the degrees of freedom associated with subsystem $B$ to obtain the reduced density matrix $\rho_A$ for subsystem $A$, i.e., 
\begin{equation}
    \rho_A = \sum_{n_{L+1},\dots,n_N} \langle n_{L+1},\dots,n_N | \rho | n_{L+1},\dots,n_N \rangle.
\end{equation}
The entanglement entropy between $A$ and $B$ is then defined as the von Neumann entropy of $\rho_A$:
\begin{equation}
S_{\mathrm{EE}} = - \mathrm{Tr} \left( \rho_A \log \rho_A \right).
\label{eq:supp-See}
\end{equation}
Physically, the entanglement entropy serves as a sensitive probe of quantum correlations and is widely used to diagnose localization in many-body systems \cite{kawabata2023entanglement,lee2024entanglement,li2023disorder,li2024emergent}. In localized phases, the entanglement entropy remains small and typically obeys area-law scaling, reflecting the limited spatial extent of quantum correlations. In contrast, delocalized or thermal phases exhibit volume-law scaling, indicative of extensive entanglement across the system. Therefore, entanglement entropy may serve as an indicator of the degree of localization in quantum many-body states.

\subsection{Inverse participation ratio (IPR)}
To quantify localization directly in the many-body Hilbert space, we expand the eigenstate in the full Fock basis $\{\ket{\bm n}\}=\{\ket{n_1, n_2,...,n_{i},..., n_{N-1}, n_N}\} $:
\begin{equation}
\ket{\psi}=\sum_{\bm n}\psi_{\bm n}\ket{\bm n},\qquad
\sum_{\bm n}|\psi_{\bm n}|^2=1,
\end{equation}
and define the inverse participation ratio
\begin{equation}
\mathrm{IPR}=\sum_{\bm n}|\psi_{\bm n}|^4.
\label{eq:supp-IPR}
\end{equation}
A basis-localized state has $\mathrm{IPR}=1$, while a state uniformly spread over $\mathcal D$ basis states has $\mathrm{IPR}_i\sim 1/\mathcal D$. Thus, decreasing IPR indicates increasing delocalization in Hilbert space.

\subsection{Real-to-complex transition and eigenstate delocalization}

\begin{figure}[H]
    \centering
    \includegraphics[width=0.9\linewidth]{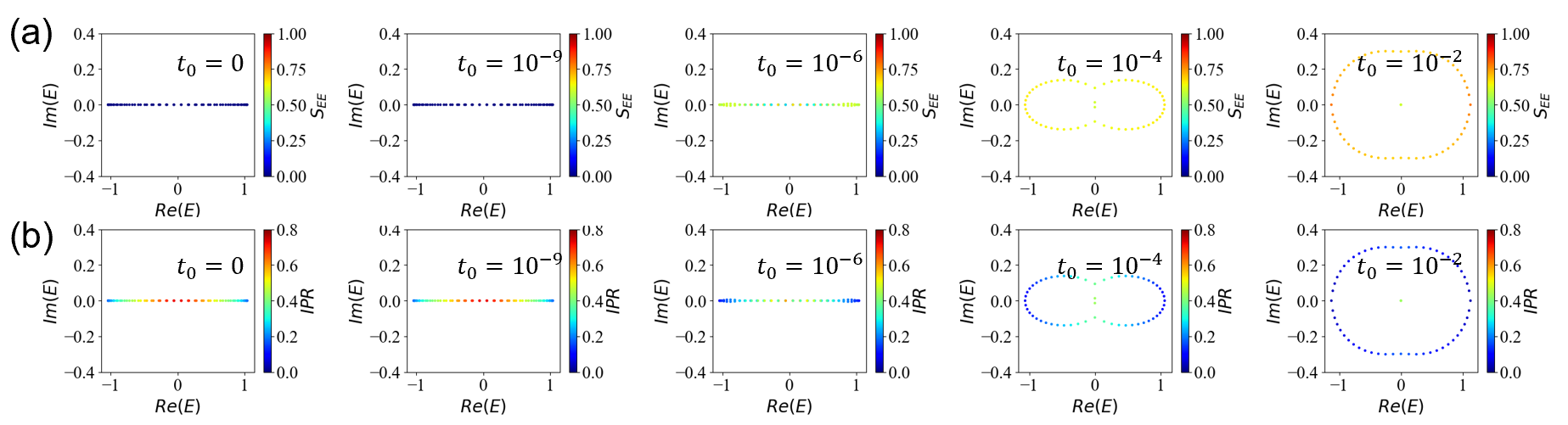}
    \caption{Eigenspectra of the model in \cref{eq:supp-model}, colored by the entanglement entropy $S_{EE}$ \cref{eq:supp-See} and the inverse participation ratio (IPR) \cref{eq:supp-IPR} for a two-particle ($n=2$) system. 
    \textbf{(a)} $S_{EE}$ of eigenstates as a function of the coupling strength $t_0$.
    \textbf{(b)} IPR of eigenstates as a function of $t_0$.
    Parameters are $t_1=0.58$, $\gamma=0.25$, and $N=30$, with OBCs imposed throughout.
    }
    \label{fig:supp-2-particles-eigs-EE-IPR}
\end{figure}

\begin{figure}[H]
    \centering
    \includegraphics[width=0.9\linewidth]{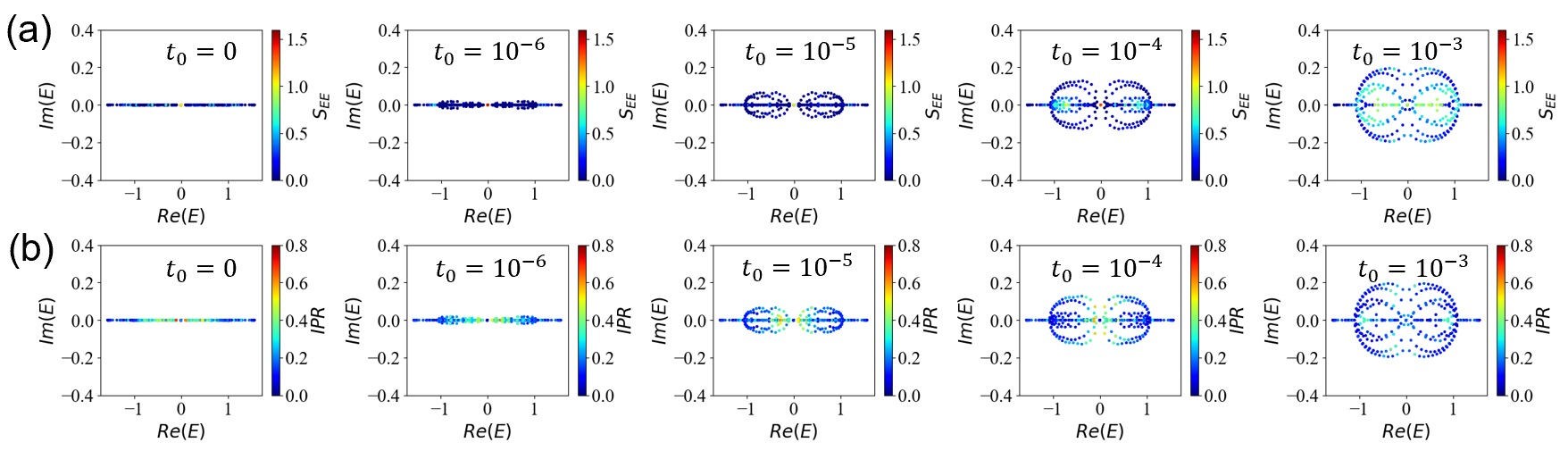}
    \caption{Eigenspectra of the model in \cref{eq:supp-model}, colored by the entanglement entropy $S_{EE}$ \cref{eq:supp-See} and the inverse participation ratio (IPR) \cref{eq:supp-IPR} for a three-particle ($n=3$) system with a maximum on-site occupation $n_{\mathrm{occ}}^{\max}=2$.
    \textbf{(a)} $S_{EE}$ of eigenstates as a function of the coupling $t_0$.
    \textbf{(b)} IPR of eigenstates as a function of the coupling $t_0$.
    Parameters are $t_1=0.58$, $\gamma=0.25$, and $N=20$, with OBCs imposed throughout.
    Note that the color scale for $S_{EE}$ used here differs from that in \cref{fig:supp-2-particles-eigs-EE-IPR} and is chosen to highlight the relative differences within the three-particle system.
    }
    \label{fig:supp-3-particle-eigs-EE-IPR}
\end{figure}

\cref{fig:supp-2-particles-eigs-EE-IPR,fig:supp-3-particle-eigs-EE-IPR} present the spectra for $n=2$ and $n=3$ bosons respectively, colored by $S_{\mathrm{EE}}$ [(a)] or $\mathrm{IPR}$ [(b)]. 
Within each figure, the columns correspond to increasing $t_0$ (values indicated in the panels), so the left-to-right progression directly tracks how eigenvalues and eigenvectors evolve as the coupling is turned on.

\paragraph{Two particles ($n=2$, \cref{fig:supp-2-particles-eigs-EE-IPR}).}
At $t_0=0$, the spectrum is purely real, and eigenstates are weakly entangled (small $S_{\mathrm{EE}}$) and Hilbert space localized (large IPR).
For $t_0=10^{-9}$--$10^{-6}$, eigenvalues remain essentially real, but the color trends already indicate gradual delocalization: $S_{\mathrm{EE}}$ increases while IPR decreases.
At $t_0=10^{-4}$, eigenvalues move off the real axis and form clear complex loops, and these complex-branch states are systematically more delocalized—higher $S_{\mathrm{EE}}$ in (a) and lower IPR in (b)—than those remaining near $\mathrm{Im}(E)=0$.
By $t_0=10^{-2}$, the spectrum is predominantly complex and the eigenstates are broadly delocalized across the band.

\paragraph{Three particles ($n=3$) with $n_{\mathrm{occ}}^{\max}=2$ (\cref{fig:supp-3-particle-eigs-EE-IPR}).}
The same correlation holds for $n=3$.
At $t_0=0$ and $10^{-6}$ the spectrum is nearly real and the eigenstates are relatively localized (low $S_{\mathrm{EE}}$, higher IPR).
Complex loops appear already at $t_0=10^{-5}$ and are accompanied by enhanced delocalization (larger $S_{\mathrm{EE}}$ and smaller IPR on the complex branches).
Increasing $t_0$ to $10^{-4}$--$10^{-3}$ expands the complex structures and pushes the spectrum toward a regime dominated by strongly entangled, Hilbert space-delocalized eigenstates.

Overall, \cref{fig:supp-2-particles-eigs-EE-IPR,fig:supp-3-particle-eigs-EE-IPR} show that the $t_0$-driven real-to-complex spectral transition is accompanied by increasing eigenstate delocalization, captured consistently by increasing $S_{\mathrm{EE}}$ and decreasing IPR. 


\section{SVI. Boundary-condition dependence of real-to-complex transition and Hilbert space connectivity}
Comparing OBCs and PBCs is important because it isolates the role of boundary-dependent Hilbert space connectivity in scaling-induced real-to-complex transition. In particular, it distinguishes the mere presence of complex eigenvalues from their scaling-induced spectral complexification -- PBC spectra are generally complex, but do not scale appreciably with system size. 
In the main text, we showed that a size-induced real-to-complex transition under OBCs, while no analogous crossover occurs under PBCs. Here, we present additional numerical results and provide an intuitive explanation, based on the structure of the Hilbert space connectivity graph, elucidating why the model in~[\cref{eq:supp-model}] does not undergo size-induced real-to-complex transition under PBCs.

\begin{figure}[H]
    \centering
    \includegraphics[width=0.9\linewidth]{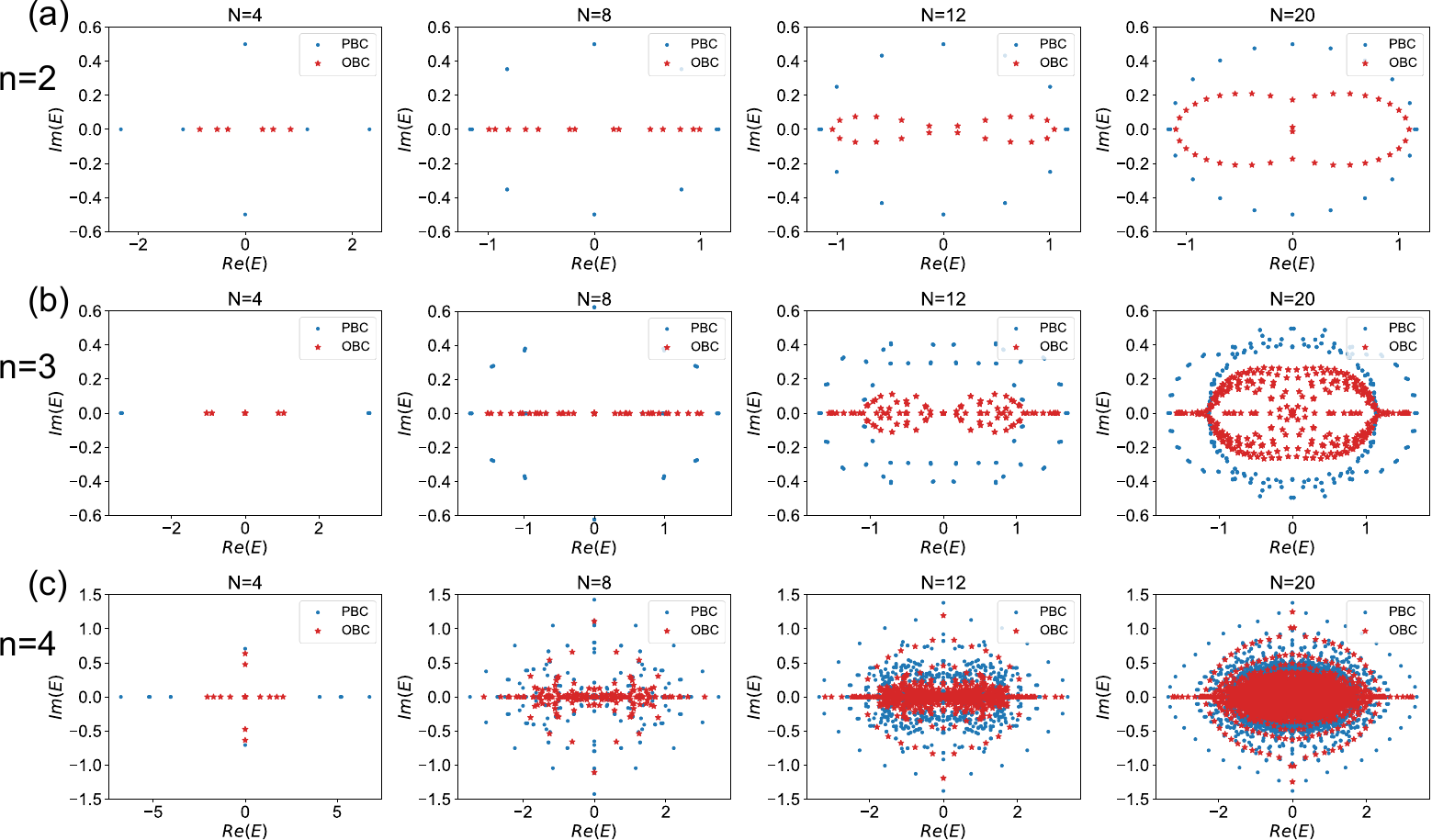}
    \caption{Eigenspectra of the model [\cref{eq:supp-model}] for various particle numbers $n$ and system sizes $N$, with a maximum on-site occupancy $n_{\mathrm{occ}}^{\max}=2$, under PBCs and OBCs.
    The parameters are set to $t_0=10^{-2}, t_1=0.58$, and $\gamma=0.25$. }
    \label{fig:supp-spectra-OBC-PBC}
\end{figure}

\cref{fig:supp-spectra-OBC-PBC} compares the eigenspectra for various particle numbers $n$ and system sizes $N$ under OBCs and PBCs. In contrast to OBCs, the spectra under PBCs show no pronounced dependence on $N$. 

\begin{figure}[htbp]
    \centering
    \includegraphics[width=0.9\linewidth]{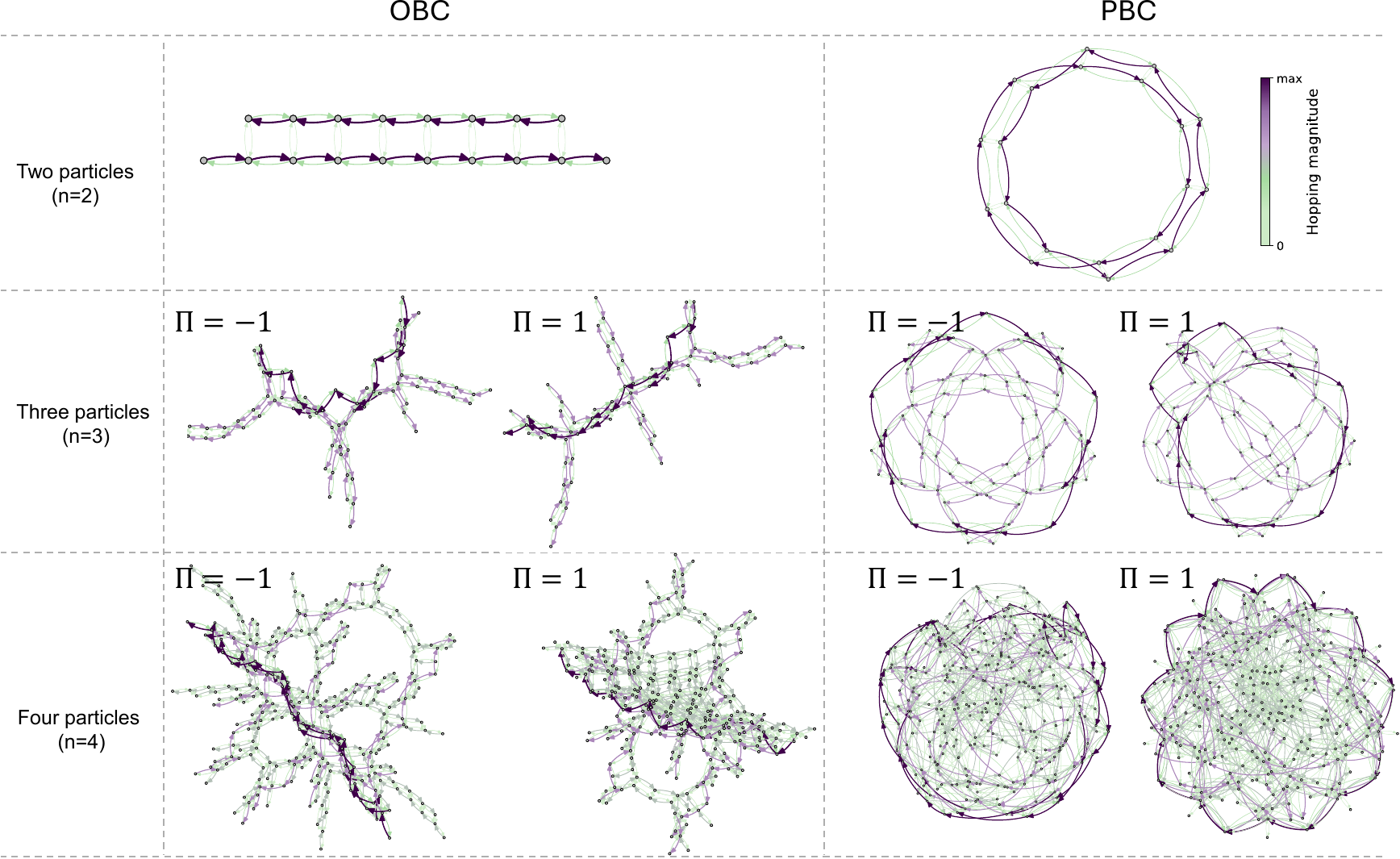}
    \caption{Comparison of the Hilbert space connectivity graphs for the model defined in \cref{eq:supp-model} under OBCs and PBCs, shown for systems with two, three, and four particles. 
    In contrast to OBCs, PBCs close chain-like connected components into loops, thereby generating numerous cyclic paths that support complex eigenvalue spectra.
    For $n>2$, this symmetry splits
    the Hilbert space into two disconnected sectors with $\Pi=\pm1$. Parameters are set to the system size $N = 10$, $t_0 = 10^{-2}$, $t_1 = 0.58$, $\gamma = 0.25$, and a maximum local occupancy of $n_{\mathrm{occ}}^{\max} = 2$.}
    \label{fig:supp-graph-OBC-PBC}
\end{figure}

We visualize the Hamiltonian as a connectivity graph in the many-body Hilbert space: each node represents an occupation configuration (Fock basis), and a directed link indicates a nonzero off-diagonal matrix element connecting two configurations (with the color/weight encoding the hopping magnitude). \cref{fig:supp-graph-OBC-PBC} compares these graphs under OBCs and PBCs at fixed $N=10$. Under OBCs (left column), the graph predominantly decomposes into coupled, chain-like components with a strong directional bias, mirroring the non-reciprocal hopping and the associated NHSE-induced drift in configuration space. As discussed in the main text, this type of directed, quasi-one-dimensional connectivity provides a minimal setting for the emergence of size-dependent real-to-complex transition.

Under PBCs (right column), the periodic identification closes these chains into loops and produces numerous cyclic paths already at the level of Hilbert space connectivity. The resulting loop-rich, strongly connected graph generically supports complex eigenvalues, which explains why the PBC spectra do not exhibit a pronounced size-driven complexification.
Thus, although PBCs can produce broadly complex spectra, their complex structure is already established at small system sizes and changes only weakly with $N$. This demonstrates that complex eigenvalues alone do not imply scaling-induced real-to-complex transition.

Moreover, because particles hop only in pairs, the model [\cref{eq:supp-model}] conserves the {$\mathbb{Z}_2$}  total position parity
$\Pi = (-1)^{\sum_i i\,n_i}$, where {$n_i=c_i^\dagger c_i$} tracks the particle number at site {$i$}. 
{For $n>2$}, this symmetry splits
the Hilbert space into two disconnected sectors with $\Pi=\pm1$. Since the accumulation behavior is identical in both parity sectors, we show only the {$\Pi=-1$} sector without loss of generality.

Taken together, the spectral and connectivity-graph comparisons show that boundary conditions qualitatively reorganize the many-body Hilbert space pathways. Open, directionally biased chain-like components produce the pronounced size-dependent complexification under OBCs, whereas their closure into loop-rich structures under PBCs removes this crossover. This establishes boundary-dependent Hilbert space connectivity as a central control mechanism for the anomalous spectral scaling.

\section{SVII. Constraint-induced Hilbert space pumping and unipolar/asymmetric bipolar localization.}
We now provide detailed supporting evidence for the second central result of the main text: the occupancy constraint restructures
the Hilbert space pathways and produces unipolar and asymmetric bipolar localization.

\begin{figure}[htbp]
    \centering
    \includegraphics[width=\linewidth]{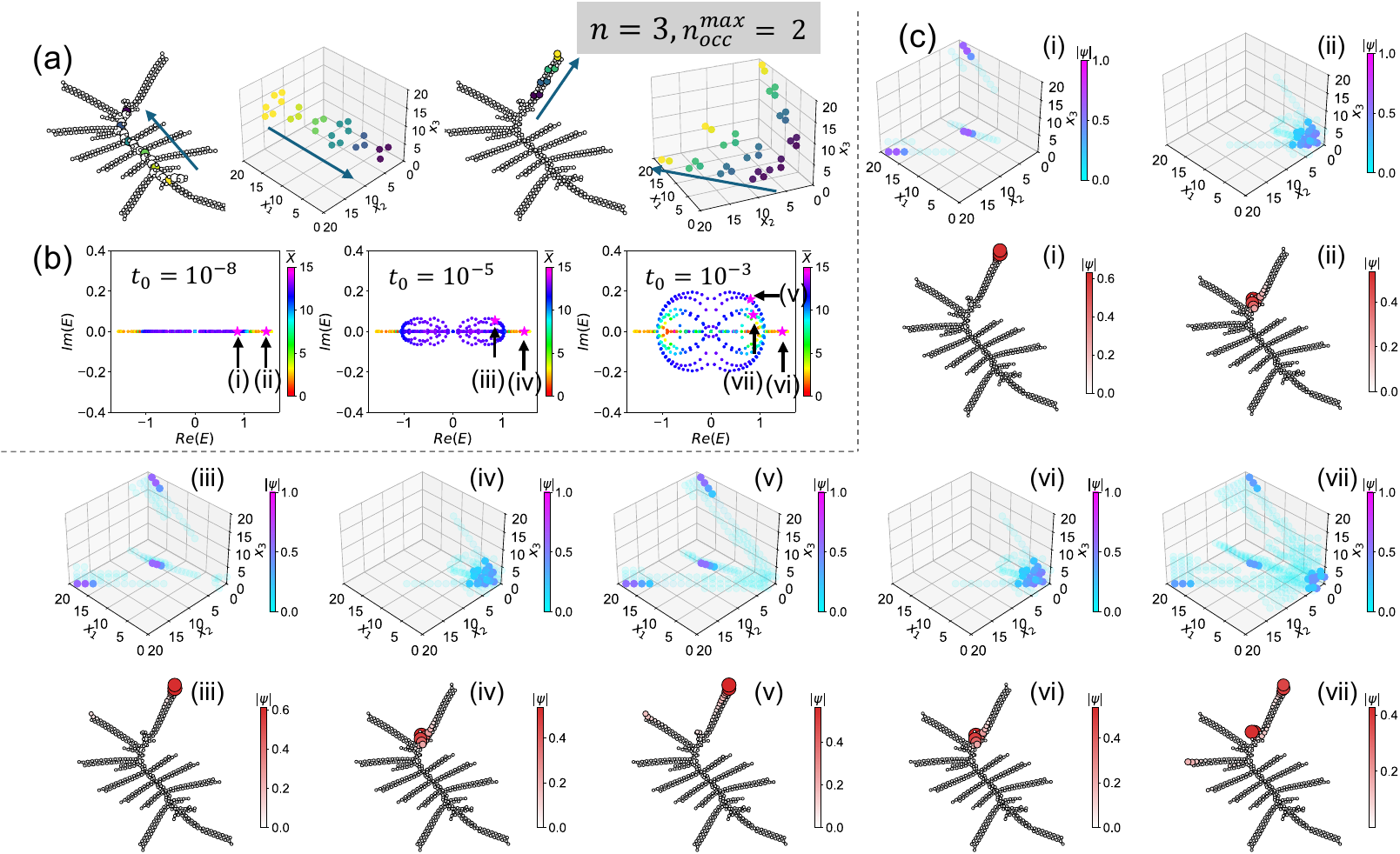}
    \caption{Unipolar and asymmetric bipolar  localization of three-particle eigenstates under OBCs with an occupancy
    constraint $n_{\mathrm{occ}}^{\max}=2$ 
    \textbf{(a)} shows representative nodes in the Hilbert space connectivity graph in terms of their corresponding 3-particle spatial coordinates $(x_1,x_2,x_3)$, where $x_i$ denotes the position of $i-$th particle in the 1D system $1\leq x_i\leq N$. 
    \textbf{(b)} Eigenspectra color-coded by the average CM position $\overline{X}$ of eigenstates [\cref{eq:supp-average-CM-position}].
    \textbf{(c)} Several representative eigenstates from (b) are illustrated as density distribution plots in three-particle configuration space $(x_1, x_2, x_3)$ (upper row) and Hilbert space connectivity graph (bottom row). 
    The parameters are set as follows: system size $N = 20$, $t_0=10^{-3}$ (unless otherwise specified), $t_1=0.58$, and $\gamma=0.25$. 
    }
    \label{fig:supp-anatomy-eigenstate}
\end{figure}

In Fig. 2 of the main text, we present the occupancy-constraint–induced unipolar and asymmetric bipolar localization of eigenstates for the three-particle case. In this section, we provide more numerical results to demonstrate such localization behavior of eigenstates by examining their profiles in both the Hilbert space connectivity graph and the particle configuration space [\cref{fig:supp-anatomy-eigenstate}]—the space spanned by all possible particle configurations $(x_1, x_2, x_3)$, where $x_i$ denotes the position of the $i$-th particle on the real-space lattice. We note that, because the particles are indistinguishable, permutations of $(x_1, x_2, x_3)$ correspond to the same physical configuration. For comparison, we also include results for systems \emph{without} the occupancy constraint, as shown in \cref{fig:supp-anatomy-eigenstate-nolimit}, which further supports that the constraint significantly enhances unipolar/bipolar localization. 

\noindent\textbf{Average center-of-mass position as a real-space diagnostic}
To characterize the spatial structure of eigenstates, we employ a real-space measure---\textit{average center-of-mass (CM) position } ($\overline{X}$) to quantify the degree of localization and spatial extent. 

The \textit{average CM position}, $\overline{X}$, of three particles in an eigenstate $\psi_n$ is defined as
\begin{equation}
\overline{X} = \sum_k \frac{x_1^{(k)} + x_2^{(k)} + x_3^{(k)}}{3} \cdot|\psi_n^{(k)}|^2,
\label{eq:supp-average-CM-position}
\end{equation}
where $x_i^{(k)}$ denotes the position of the $i$-th particle in the $k$-th Fock basis configuration, and $|\psi_n^{(k)}|^2$ is the probability weight of that configuration. 
Importantly, $\overline{X}$ is not meant to replace a full localization measure; instead it provides a
\emph{direct and physically transparent label} for \emph{where} an eigenstate predominantly lives in real space.
This makes it ideal for color-coding the eigenspectrum:
\begin{itemize}[noitemsep, topsep=0pt]
\item For a \textbf{unipolar} state localized near the left edge (e.g. configurations close to $(1,1,1)$),
one expects $\overline{X}\approx 1$.
\item For an \textbf{asymmetric bipolar} state localized near $(1,N,N)$, the CM shifts rightward, giving
$\overline{X}\approx 2N/3\approx 13$.
\item For \textbf{hybrid/mixed} states with appreciable weight in both patterns, $\overline{X}$ takes intermediate values.
\end{itemize}
Therefore, if the spectrum colored by $\overline{X}$ shows well-separated color regions, it directly signals that
different spectral windows are dominated by distinct eigenstate localization patterns.

\noindent\textbf{Unipolar and asymmetric bipolar localization due to occupancy constraint}
In \cref{fig:supp-anatomy-eigenstate}(a), we illustrate the correspondence between nodes in the Hilbert space connectivity graph (left) and their 3-particle spatial coordinates (right). The connectivity graph consists of a central backbone and several branches that extend outward from it. Notably, the backbone—connecting multiple branches-corresponds to basis states where all three particles are clustered together, i.e., $x_1 \approx x_2 \approx x_3$. In contrast, the top branch of the coupled-chain structure corresponds to states where two particles form a bound pair, $x_2 \approx x_3$, while the third remains fixed at $x_1 = 1$, i.e., configurations characterized by $x_1 = 1$, $x_2 \approx x_3$. 

In \cref{fig:supp-anatomy-eigenstate}(b), we present the eigenspectra of the three-particle system subject to a maximum on-site occupancy constraint $n_{\mathrm{occ}}^{\max} = 2$, color-coded by $\overline{X}$ of eigenstates [\cref{eq:supp-average-CM-position}]. Results are shown for coupling strengths $t_0 = 10^{-8},\ 10^{-5},\ 10^{-3}$. For small $t_0$, the spectrum separates into regions with roughly two $\overline{X}$ values [\cref{fig:supp-anatomy-eigenstate}(b); compare (i,iii) with (ii,iv) in the upper row],
indicating two distinct real-space accumulation patterns. It is confirmed in \cref{fig:supp-anatomy-eigenstate}(c): eigenstates taken from the ``small-$\overline{X}$'' spectral region exhibit unipolar accumulation near $(1,1,1)$,
whereas those from the ``large-$\overline{X}$'' region exhibit asymmetric bipolar accumulation near
$(1,N,N)$. As $t_0$ increases, other than the existing two patterns [\cref{fig:supp-anatomy-eigenstate}(b-c)(v, vi), upper row], both start to hybridize, producing intermediate-$\overline{X}$ eigenstates with mixed unipolar/bipolar
character (see, e.g., \cref{fig:supp-anatomy-eigenstate}(b-c)(vii), upper row).
Consistently, the distribution of eigenstates on the Hilbert space connectivity graph separates into two classes—one localized near the top chain [\cref{fig:supp-anatomy-eigenstate}(c)(i,iii), bottom row] and the other near the end of the backbone [\cref{fig:supp-anatomy-eigenstate}(c)(ii,iv), bottom row]—which begin to mix as $t_0$ increases [\cref{fig:supp-anatomy-eigenstate}(c)(vii), bottom row].

We next remove the occupancy constraint (allowing triple occupancy), and re-examine the eigenstates in
\cref{fig:supp-anatomy-eigenstate-nolimit}. In this case, the $\overline{X}$-colored spectrum becomes much less segregated [\cref{fig:supp-anatomy-eigenstate-nolimit}(a)], and representative eigenstates show
substantially weaker unipolar/bipolar accumulation [\cref{fig:supp-anatomy-eigenstate-nolimit}(b)]. This comparison emphasizes that the occupancy constraint is not a minor quantitative detail.

\begin{figure}
    \centering
    \includegraphics[width=0.6\linewidth]{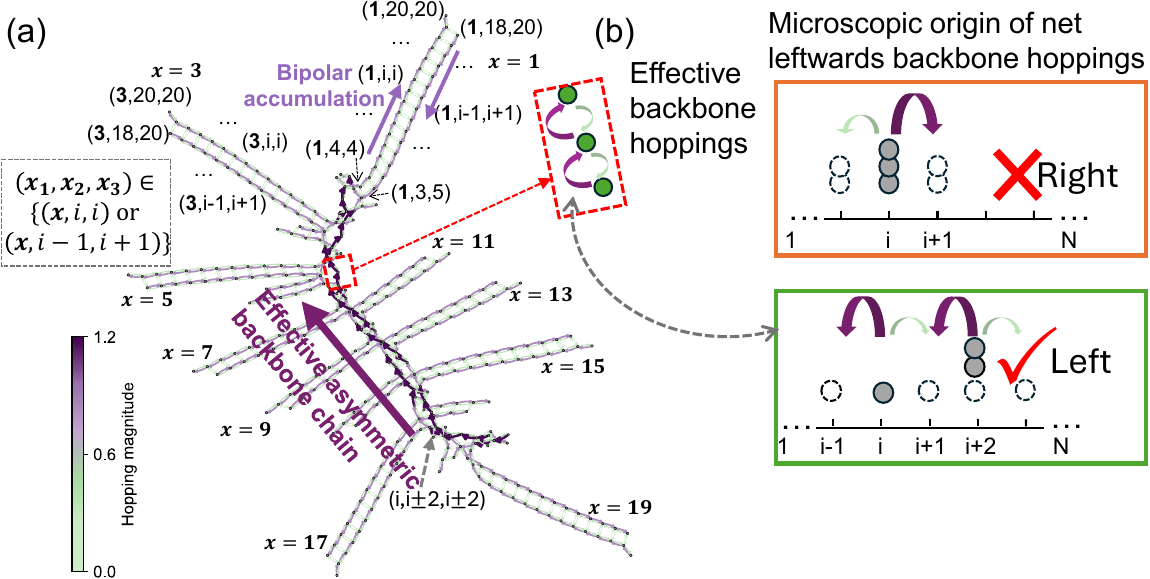}
    \caption{Mechanism of unipolar and asymmetric bipolar localization of eigenstates in a three-particle system under OBCs with a maximum occupancy constraint $n_{\mathrm{occ}}^{\max}=2$. 
    \tb{(a)} Hilbert space connectivity graph of \cref{eq:supp-model} showing a central backbone connected to multiple branches. The purple arrow denotes the directional hopping bias along the backbone. Each node represents a configuration state labeled by $(x_1, x_2, x_3)$, where $x_i$ denotes the position of the $i$-th particle.
    \tb{(b)} Bias-enhanced hopping process arising from the bosonic operator algebra (bottom), while the opposite process is forbidden by the occupancy constraint (top).
    The parameters are set to $t_0=10^{-3}$, $t_1=0.58$, system size $N=20$, and $\gamma=0.25$.
    }
    \label{fig:supp-exp-corner-loc}
\end{figure}

\begin{figure}
    \centering
    \includegraphics[width=\linewidth]{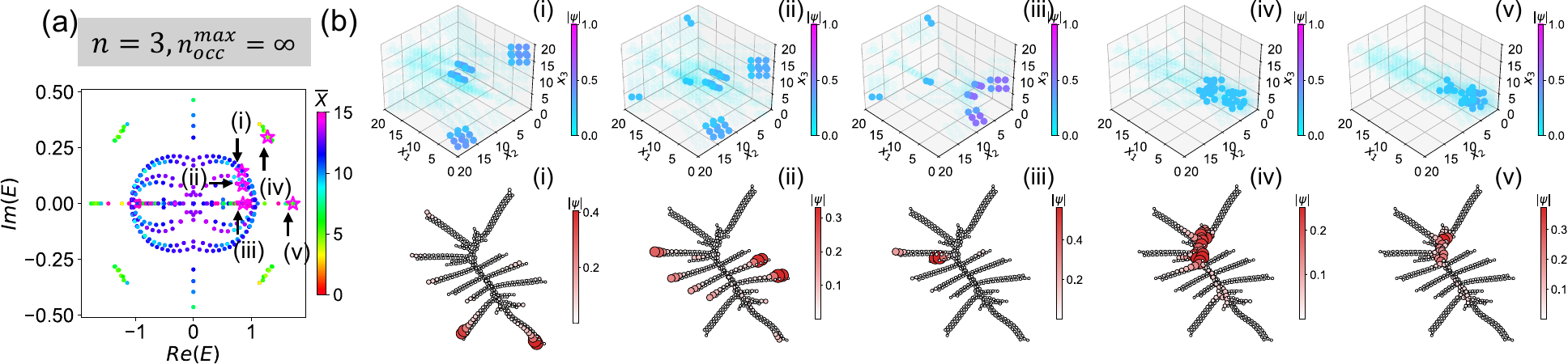}
    \caption{Weakening of unipolar and bipolar localization in a three-particle system under OBCs without the on-site occupancy constraint.
    \textbf{(a)} Eigenspectra color-coded by $\overline{X}$ of eigenstates [\cref{eq:supp-average-CM-position}]. 
    \textbf{(b)} Several representative eigenstates (pink hollow stars) are illustrated as density distribution plots in three-particle configuration space $(x_1, x_2, x_3)$ and on a Hilbert space connectivity graph. 
    The parameters used are system size $N=20$, $t_0 = 10^{-3}$, $t_1 = 0.58$, and $\gamma = 0.25$.
    }
    \label{fig:supp-anatomy-eigenstate-nolimit}
\end{figure}

Here, we briefly recapitulate the mechanism underlying constraint-induced many-body unipolar and asymmetric bipolar localization, as discussed in detail in the main text. The key ingredient is the on-site occupancy constraint ($n_{\mathrm{occ}}^{\max}=2$), which removes all triple-occupancy configurations from the many-body Hilbert space. In the corresponding Hilbert space connectivity graph, this removal deletes specific hopping paths and, crucially, can eliminate the \emph{reverse} counterpart of an otherwise bosonically enhanced process [\cref{fig:supp-exp-corner-loc}(b)]. The resulting imbalance between forward and backward transitions breaks effective reciprocity, leading to probability ``pumping'' in the backbone in the Hilbert space connectivity graph, as shown in \cref{fig:supp-exp-corner-loc}(a). This preferential hopping along the backbone [\cref{fig:supp-exp-corner-loc}(a)] channels eigenstates toward the top chain, where they accumulate at both ends—manifesting as unipolar and asymmetric bipolar localization.

As discussed before, three qualitatively distinct types of eigenstate behavior emerge in different spectral regions for large coupling $t_0$ [\cref{fig:supp-anatomy-eigenstate}(b), e.g., (v), (vi), (vii)]. Here, we verify the robustness of these behaviors by varying the system size $N$ while keeping the coupling parameter $t_0$ fixed. To eliminate size-dependent artifacts, we focus on the normalized version of the average center-of-mass position $\overline{X}/N$ [\cref{eq:supp-average-CM-position}].
\begin{figure}[h!]
    \centering
    \includegraphics[width=\linewidth]{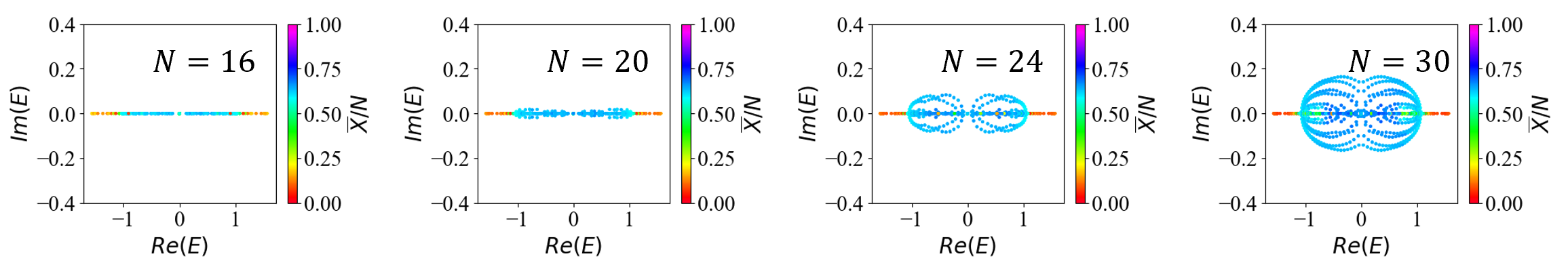}
    \caption{Eigenenergy spectra of the model in \cref{eq:supp-model}, colored by the normalized mean center-of-mass (CM) position $\overline{X}/N$ defined in \cref{eq:supp-average-CM-position}, for a three-particle system with OBCs at various system sizes $N$, with a maximum on-site occupation $n_{\mathrm{occ}}^{\max}=2$.
    The parameters are set to $t_0=10^{-6}$, $t_1=0.58$, and $\gamma=0.25$.
    }
    \label{fig:supp-spectra-quantity-N}
\end{figure}
\newpage

\section{SVIII. Proposal for measuring the scaling-induced maximum imaginary eigenenergy ($\text{maxIm}(E)$) on a quantum processor}
In this section, we describe how to employ rapidly advancing digital quantum simulators to probe the transition from real to complex eigenenergies [Fig.~1(c) in the main text], triggered by varying either the coupling $ t_0 $ or the system size $ N $ in few-body non-Hermitian models ($n\le3$). Our goal is to extract the largest imaginary component of the spectrum by monitoring long-time dynamics.

\noindent\textbf{Qubit encoding} We begin by encoding the system's basis states into qubit states. Consider a system with $n$ indistinguishable particles on $N$ lattice sites. A basis state $\ket{x_1, \dots,x_i,\dots x_n}$ specifies the positions of particles, where each $x_i \in [1, N]$. These many-body configurations are then mapped to computational-basis states of $N_q$ qubits in lexicographic ascending order. For systems without an occupation limit, the size of the Hilbert space is $\binom{N-1+n}{n}$, and the number of qubits required is $N_q = \log_2 \binom{N-1+n}{n} \sim n\log_2 N$. For systems with a maximum occupation number $n_{\mathrm{occ}}^{\max} = 2$ per site, the Hilbert space dimension becomes $\log_2 \sum_{j=0}^{\lfloor n/2\rfloor} \binom{N}{j}\binom{N-j}{n-2j}$ and the number of qubits scales as $\log_2 \sum_{j=0}^{\lfloor n/2 \rfloor} \binom{N}{j} \binom{N-j}{n-2j} \sim n\log N$. 

\noindent\textbf{Trotterization}
Any Hamiltonian $\mathcal{H}$ (possibly non-Hermitian) can be decomposed as $\mathcal{H}=\mathcal{H}_{\mathrm{H}}-i\mathcal{H}_{A}$, where $\mathcal{H}_{\mathrm{H}}$ is Hermitian and $-i\mathcal{H}_{A}$ is anti-Hermitian~\cite{shen2026simulating}. 
We approximate the time evolution operator $U(t)=e^{-i\mathcal{H}t}$ by Trotterizing it into $m$ steps: 
\begin{equation}
    \ket{\psi(t)} = U(t) \ket{\psi(0)} = \left(U_{\Delta t}^{\mathrm{H}} U_{\Delta t}^{A}\right)^m \ket{\psi(0)} + \mathcal{O}(t\Delta t), 
    \label{eq:supp-timevo}
\end{equation} where $t = m \Delta t$, $U_{\Delta t}^{\mathrm{H}} = e^{-i\mathcal{H}_{\mathrm{H}}\Delta t}$ and $U_{\Delta t}^{\mathrm{A}} = e^{-\mathcal{H}_{\mathrm{A}}\Delta t}$.
The unitary piece $U_{\Delta t}^{\mathrm{H}}$ can be implemented using the first-order Trotter-Lie product formula~\cite{Hatano2005finding}, which rewrites $\mathcal{H}_{\mathrm{H}}$ in the Pauli basis. The time evolution operator is expressed as 
\begin{equation}
    U_{\Delta t}^H = \prod_{k=1}^{k_H} e^{-i\alpha_k \sigma^k \Delta t},
    \label{eq:supp-timevo-H}
\end{equation} 
where each $\sigma^k$ is a Pauli string and $\alpha_k \in \mathbb{R}$. These Pauli rotations can be readily executed on quantum circuits.

\noindent\textbf{LCU approximation for non-unitary evolution} 
The remaining challenge lies in simulating the non-unitary evolution $U_{\Delta t}^{A} = e^{-\mathcal{H}_{A} \Delta t}$ in \cref{eq:supp-timevo}. A gate-based quantum processor
natively implements unitary transformations, and therefore $U_{\Delta t}^{A}$ cannot be implemented deterministically by a closed quantum circuit acting only on the system qubits. We instead embed this non-unitary operator into a larger unitary circuit through a linear combination of unitaries (LCU)~\cite{childs2012hamiltonian,berry2015simulating,koh2025interacting,shen2026simulating}.

Although $\mathcal{H}_{A}$ is Hermitian, it need not be positive
semidefinite. Let $\lambda_{\min}(\mathcal{H}_{A})$ denote its lowest
eigenvalue within the Hilbert space sector being simulated. We
introduce the uniform spectral shift
\begin{equation}
    \widetilde{\mathcal{H}}_{A}
    =
    \mathcal{H}_{A}
    +
    \lambda\mathbb{I},
    \qquad
    \lambda
    \geq
    \max\left\{
        0,
        -\lambda_{\min}(\mathcal{H}_{A})
    \right\},
    \label{eq:supp-HA-shift}
\end{equation}
such that $\widetilde{\mathcal{H}}_{A}\succeq0$. We denote normalization to unit trace by
$\mathcal{N}[X]
\equiv
\frac{X}{\operatorname{Tr}[X]}.$
This uniform shift does not
change the normalized state because 
\begin{equation}
\begin{aligned}
    \mathcal{N}\!\left[
        e^{-\widetilde{\mathcal{H}}_{A}\Delta t}
        \rho
        e^{-\widetilde{\mathcal{H}}_{A}\Delta t}
    \right]
    &=
    \mathcal{N}\!\left[
        e^{-2\lambda\Delta t}
        e^{-\mathcal{H}_{A}\Delta t}
        \rho
        e^{-\mathcal{H}_{A}\Delta t}
    \right]
    \\
    &=
    \mathcal{N}\!\left[
        e^{-\mathcal{H}_{A}\Delta t}
        \rho
        e^{-\mathcal{H}_{A}\Delta t}
    \right].
\end{aligned}
\label{eq:supp-shift-normalization}
\end{equation}
We may therefore construct the circuit using
\begin{equation}
    R
    =
    \sqrt{\widetilde{\mathcal{H}}_{A}},
    \qquad
    R^{2}
    =
    \widetilde{\mathcal{H}}_{A},
    \label{eq:supp-R-def}
\end{equation}
where $R$ is Hermitian and positive semidefinite.

Define
\begin{equation}
    \theta
    =
    \sqrt{2\Delta t},
    \qquad
    U_{\pm}
    =
    e^{\pm iR\theta}.
    \label{eq:supp-U-pm}
\end{equation}
Since $R$ is Hermitian, both $U_{+}$ and $U_{-}$ are unitary. Their
equal coherent sum gives
\begin{equation}
\begin{aligned}
    W_{\Delta t}
    &\equiv
    \frac{1}{2}
    \left(
        U_{+}+U_{-}
    \right)
    =
    \cos\left(
        R\sqrt{2\Delta t}
    \right)
    \\
    &=
    \mathbb{I}
    -
    \widetilde{\mathcal{H}}_{A}\Delta t
    +
    \frac{1}{6}
    \widetilde{\mathcal{H}}_{A}^{\,2}
    \Delta t^{2}
    +
    \mathcal{O}(\Delta t^{3})
    \\
    &=
    e^{-\widetilde{\mathcal{H}}_{A}\Delta t}
    -
    \frac{1}{3}
    \widetilde{\mathcal{H}}_{A}^{\,2}
    \Delta t^{2}
    +
    \mathcal{O}(\Delta t^{3}).
\end{aligned}
\label{eq:supp-timevo-A}
\end{equation}
Thus, $W_{\Delta t}$ reproduces the desired imaginary-time propagator
with a local error of order $\mathcal{O}(\Delta t^{2})$. Equivalently,
for an eigenstate $R\ket{r}=r\ket{r}$, the two unitary branches
acquire phases $e^{\pm ir\sqrt{2\Delta t}}$, whose coherent sum gives
\begin{equation}
    W_{\Delta t}\ket{r}
    =
    \cos\left(
        r\sqrt{2\Delta t}
    \right)\ket{r}
    =
    \left[
        1-r^{2}\Delta t
        +
        \mathcal{O}(\Delta t^{2})
    \right]\ket{r}.
\end{equation}
The non-unitary attenuation therefore originates from interference
between the two unitary branches.

The coherent sum in \cref{eq:supp-timevo-A} is implemented using one
ancillary qubit. Starting from $\ket{0}_{a}\ket{\psi}$, a Hadamard
gate prepares the ancilla in
$\ket{+}_{a}=(\ket{0}_{a}+\ket{1}_{a})/\sqrt{2}$. We then apply
\begin{equation}
    \operatorname{SELECT}(U)
    =
    \ket{0}\!\bra{0}_{a}
    \otimes
    U_{-}
    +
    \ket{1}\!\bra{1}_{a}
    \otimes
    U_{+}.
    \label{eq:supp-select-U}
\end{equation}
After a second Hadamard gate on the ancilla, the joint state is
\begin{equation}
\begin{aligned}
    &\left(
        H_{a}\otimes\mathbb{I}
    \right)
    \operatorname{SELECT}(U)
    \left(
        H_{a}\otimes\mathbb{I}
    \right)
    \ket{0}_{a}\ket{\psi}
    \\
    &\qquad =
    \ket{0}_{a}
    \frac{U_{-}+U_{+}}{2}
    \ket{\psi}
    +
    \ket{1}_{a}
    \frac{U_{-}-U_{+}}{2}
    \ket{\psi}
    \\
    &\qquad =
    \ket{0}_{a}
    W_{\Delta t}\ket{\psi}
    +
    \ket{1}_{a}
    W_{\Delta t}^{\perp}\ket{\psi},
\end{aligned}
\label{eq:supp-LCU-state}
\end{equation}
where
\begin{equation}
    W_{\Delta t}^{\perp}
    =
    \frac{U_{-}-U_{+}}{2}
    =
    -i\sin\left(
        R\sqrt{2\Delta t}
    \right)
\end{equation}
is the complementary branch.

The ancilla is measured in the computational basis, equivalently in
the $\sigma^{x}$ basis before the second Hadamard gate. Conditional
on obtaining the outcome $0$, the LCU step succeeds and realizes
\begin{equation}
\begin{aligned}
    \mathcal{M}_{A}^{\mathrm{LCU}}(\rho)
    &=
    \frac{
        W_{\Delta t}
        \rho
        W_{\Delta t}^{\dagger}
    }{
        \operatorname{Tr}\!\left[
            W_{\Delta t}
            \rho
            W_{\Delta t}^{\dagger}
        \right]
    }
    \\
    &=
    \mathcal{N}\!\left[
        e^{-\mathcal{H}_{A}\Delta t}
        \rho
        e^{-\mathcal{H}_{A}\Delta t}
    \right]
    +
    \mathcal{O}(\Delta t^{2}).
\end{aligned}
\label{eq:supp-MA-map}
\end{equation}

Finally, the shift in \cref{eq:supp-HA-shift} must be removed when
reporting spectral quantities. In particular,
\begin{equation}
    \operatorname{Im}(E)
    =
    -\left\langle
        \mathcal{H}_{A}
    \right\rangle
    =
    -\left\langle
        \widetilde{\mathcal{H}}_{A}
    \right\rangle
    +
    \lambda.
    \label{eq:supp-ImE-shift}
\end{equation}
The shift therefore leaves the normalized time-evolved state and the
selected eigenstate unchanged, but its known offset must be restored
in the inferred imaginary eigenenergy.

\noindent\textbf{Extracting the max imaginary energy} To detect the spectral transition, we focus on measuring the maximum imaginary part of the eigenenergies, $\max \operatorname{Im}(E)$. As the system size increases, the spectrum transitions from real to complex. One can access the dominant eigenvalue with the largest $\operatorname{Im}(E)$ by analyzing the long-time evolution of an arbitrary initial state. In the asymptotic limit, this evolution isolates the eigenstate with the largest imaginary component $\ket{\phi_{max}}$: 
\begin{equation} 
\begin{aligned}
\ket{\psi(t)} &\propto e^{-i\mathcal{H}t} \ket{\psi(0)} \approx e^{-i\mathrm{Re}(E)t} e^{\operatorname{Im}(E)t} \ket{\psi(0)} \\
& \xrightarrow{t\to \infty} \ket{\phi_{max}},
\end{aligned}
\end{equation} 
where $\ket{\psi(0)}$ is our initial state. The long-time evolution is implemented via the Trotterization scheme described in \cref{eq:supp-timevo}, where $U_{\Delta t}^{H}$ and $U_{\Delta t}^{A}$ are realized using \cref{eq:supp-timevo-H} and \cref{eq:supp-timevo-A}, respectively.
The imaginary part $\operatorname{Im}(E)$ can then be extracted by measuring the expectation value of the anti-Hermitian component on $| \phi_{max} \rangle$: 
\begin{equation} 
    \operatorname{Im}(E) = -\langle \phi_{max} | \mathcal{H}_A | \phi_{max} \rangle. 
\end{equation}

We measure this expectation value using the Hamiltonian averaging technique~\cite{mcclean2014exploiting, mcclean2016theory, peruzzo2014variational,koh2025interacting}, which we implement as a three-step process. First, the Hamiltonian $\mathcal{H}_A$ is decomposed into a sum of $K_A$ local Pauli strings, $\mathcal{H}_A = \sum_{k=1}^{K_A} \beta_k \chi^{(k)}$, where $\chi^{(k)}$ is a tensor product of Pauli operators (e.g., $\sigma_i^x \sigma_j^z$) and $\beta_k \in \mathbb{R}$. By linearity, the expectation value is $\langle \mathcal{H}_A \rangle = \sum_k \beta_k \langle \chi^{(k)} \rangle$.

Then, we measure the expectation value $\langle \chi^{(k)} \rangle$ for each term. A quantum processor performs measurements in a fixed computational (or $\sigma^z$) basis. To measure operators containing $\sigma^x$ or $\sigma^y$, a change of basis is required. This is achieved by appending single-qubit rotations to the circuit immediately before the physical measurement~\cite{nielsen2010quantum}. For a given qubit $j$, the expectation values of Pauli operators are measured as follows. To measure $\langle \sigma_j^z \rangle$, no rotation is applied prior to the $Z$-basis measurement. To access $\langle \sigma_j^x \rangle$, a Hadamard gate ($H$) is applied, mapping the $X$ basis to the $Z$ basis. For $\langle \sigma_j^y \rangle$, an $S^\dagger H$ rotation is applied before measurement, effectively transforming the $Y$ basis to the $Z$ basis. For a multi-qubit string $\chi^{(k)}$, these rotations are applied in parallel to all relevant qubits. The expectation value $\langle \chi^{(k)} \rangle$ is estimated from measurement statistics collected over many repeated experimental runs. To reduce the number of circuits, mutually commuting Pauli strings can be grouped and measured simultaneously \cite{tilly2022variational,crawford2021efficient}.

Finally, ${\operatorname{Im}}(E)$ is reconstructed by summing the weighted expectation values $\beta_k \langle \chi^{(k)} \rangle$. This scheme enables the characterization of the real-to-complex spectral transition on quantum hardware, providing insight into the underlying many-body non-Hermitian dynamics.

\end{document}